\documentclass{siamltex}

\usepackage{amssymb,amsmath,amsfonts,amssymb}
\usepackage{graphics,graphicx,color}

\newcommand{\eps}{\varepsilon}
\newcommand{\dsum}{\displaystyle\sum}
\newcommand{\dint}{\displaystyle\int}
\newcommand{\bx}{{\bf x}} \newcommand{\vx}{{\bf x}}
\newcommand{\vz}{{\bf z}} \newcommand{\vy}{{\bf y}}
\newcommand{\vb}{{\bf b}} \newcommand{\by}{{\bf y}}
\newcommand{\bb}{{\bf b}} \newcommand{\bz}{{\bf z}}
\newcommand{\bk}{{\bf k}} \newcommand{\bp}{{\bf p}}
 \newcommand{\bq}{{\bf q}}
\newcommand{\bu}{{\bf u}} \newcommand{\bv}{{\bf v}}
\newcommand{\vE}{{\bf E}} \newcommand{\vH}{{\bf H}}
\newcommand{\vu}{{\bf u}} \newcommand{\bw}{{\bf w}}

\renewcommand{\d}{3}
\newcommand{\vhatk}{{\hat{\bf k}}}\newcommand{\vhatp}{{\hat{\bf p}}}

\newcommand{\Bxi}{\boldsymbol{\xi}}  \newcommand{\bphi}{\boldsymbol{\phi}}

\newcommand{\bpsi}{\boldsymbol{\bpsi}}  
 \newcommand{\vxi}{\boldsymbol{\xi}}
  \def\bS{{\bf S}}
 
\newcommand{\pdr}[2]{\dfrac{\partial{#1}}{\partial{#2}}}
\newcommand{\pdrt}[2]{\dfrac{\partial^2{#1}}{\partial{#2}^2}}

\newcommand{\Rm}{\mathbb R}

\begin{document}

\title{Time Reversal for Waves in Random Media}

\author{Guillaume Bal\thanks{Department of Applied Physics and Applied
    Mathematics, Columbia University, New York, NY 10027, USA; {e-mail:
      gb2030@columbia.edu}}
    \and  Leonid Ryzhik \thanks{Department of Mathematics,
University of Chicago, Chicago, IL 60637, USA; e-mail:
ryzhik@math.uchicago.edu}}

\maketitle
\begin{abstract}
  In time reversal acoustics experiments, a signal is emitted from a
  localized source, recorded at an array of receivers-transducers,
  time reversed, and finally re-emitted into the medium.  A celebrated
  feature of time reversal experiments is that the refocusing of the
  re-emitted signals at the location of the initial source is improved
  when the medium is heterogeneous.  Contrary to intuition, multiple
  scattering enhances the spatial resolution of the refocused signal
  and allows one to beat the diffraction limit obtained in homogeneous
  media.  This paper presents a quantitative explanation of time
  reversal and other more general refocusing phenomena for general
  classical waves in heterogeneous media.  The theory is based on the
  asymptotic analysis of the Wigner transform of wave fields in the
  high frequency limit.  Numerical experiments complement the theory.
\end{abstract}

\begin{keywords}
  Waves in random media, time reversal, refocusing,
  radiative transfer equations, diffusion approximation.
\end{keywords}

\begin{AMS}
  35F10 35B40 82D30
\end{AMS}

\pagestyle{myheadings}
\thispagestyle{plain}
\markboth{G. BAL and L. RYZHIK}
  {TIME REVERSAL FOR WAVES}

\section{Introduction}
\label{sec:intro}

In time reversal experiments, acoustic waves are emitted from a
localized source, recorded in time by an array of
receivers-transducers, time reversed, and re-transmitted into the
medium, so that the signals recorded first are re-emitted last and
vice versa
\cite{DRF-PRL95,dowling-JASA92,Fink-Prada-01,HSK99,khosla-dowling-JASA01,
  KHS97}. The re-transmitted signal refocuses at the location of the
original source with a modified shape that depends on the array of
receivers. The salient feature of these time reversal experiments is
that refocusing is much better when wave propagation occurs in
complicated environments than in homogeneous media.  Time reversal
techniques with improved refocusing in heterogeneous medium have found
important applications in medicine, non-destructive testing,
underwater acoustics, and wireless communications (see the above
references). It has been also applied to imaging in weakly random
media \cite{BBPT,Fink-Prada-01}.

\begin{figure}[htbp]
  \begin{center}
    \includegraphics[height=5cm]{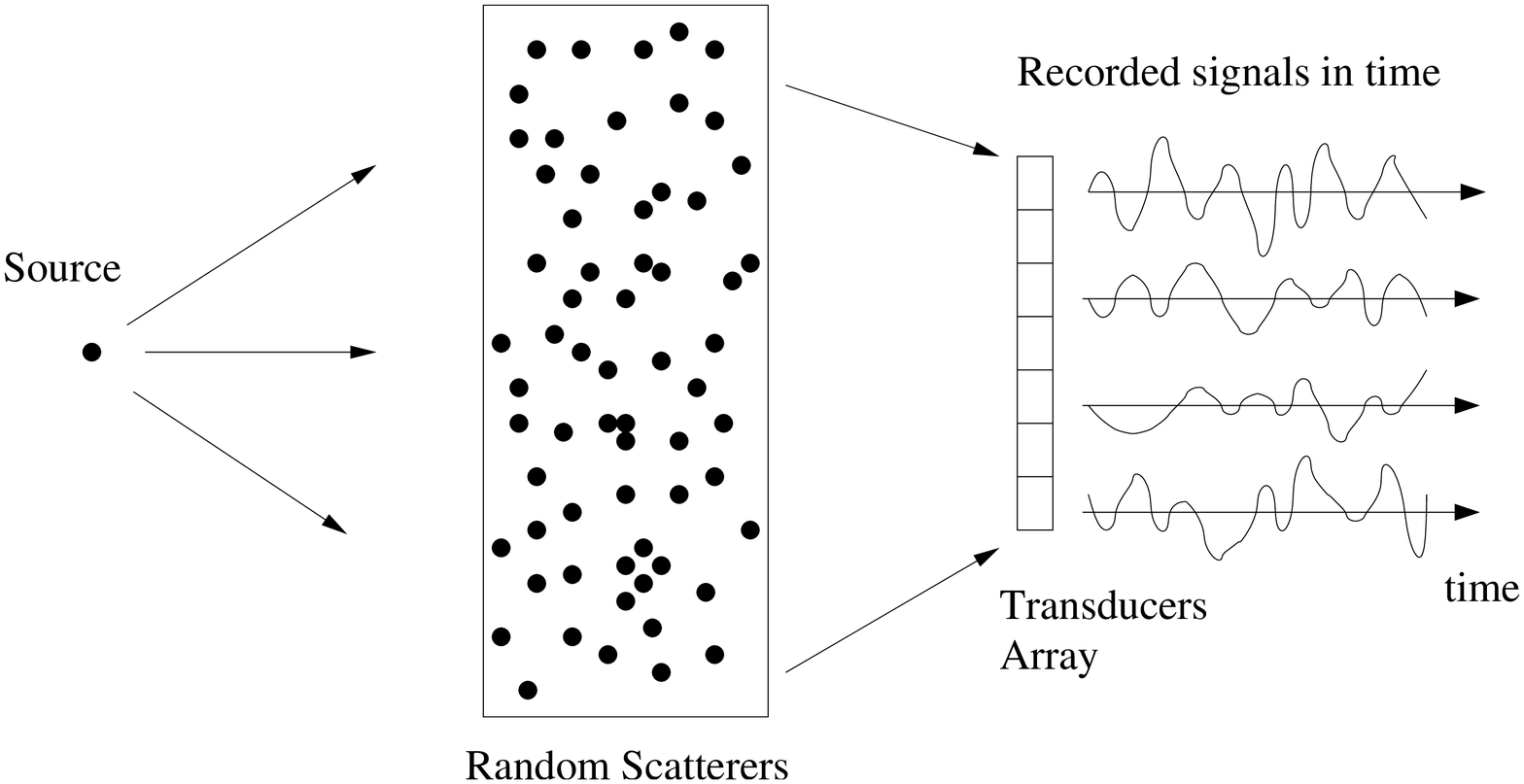} \\
    $\mbox{}$ \\
    \includegraphics[height=5cm]{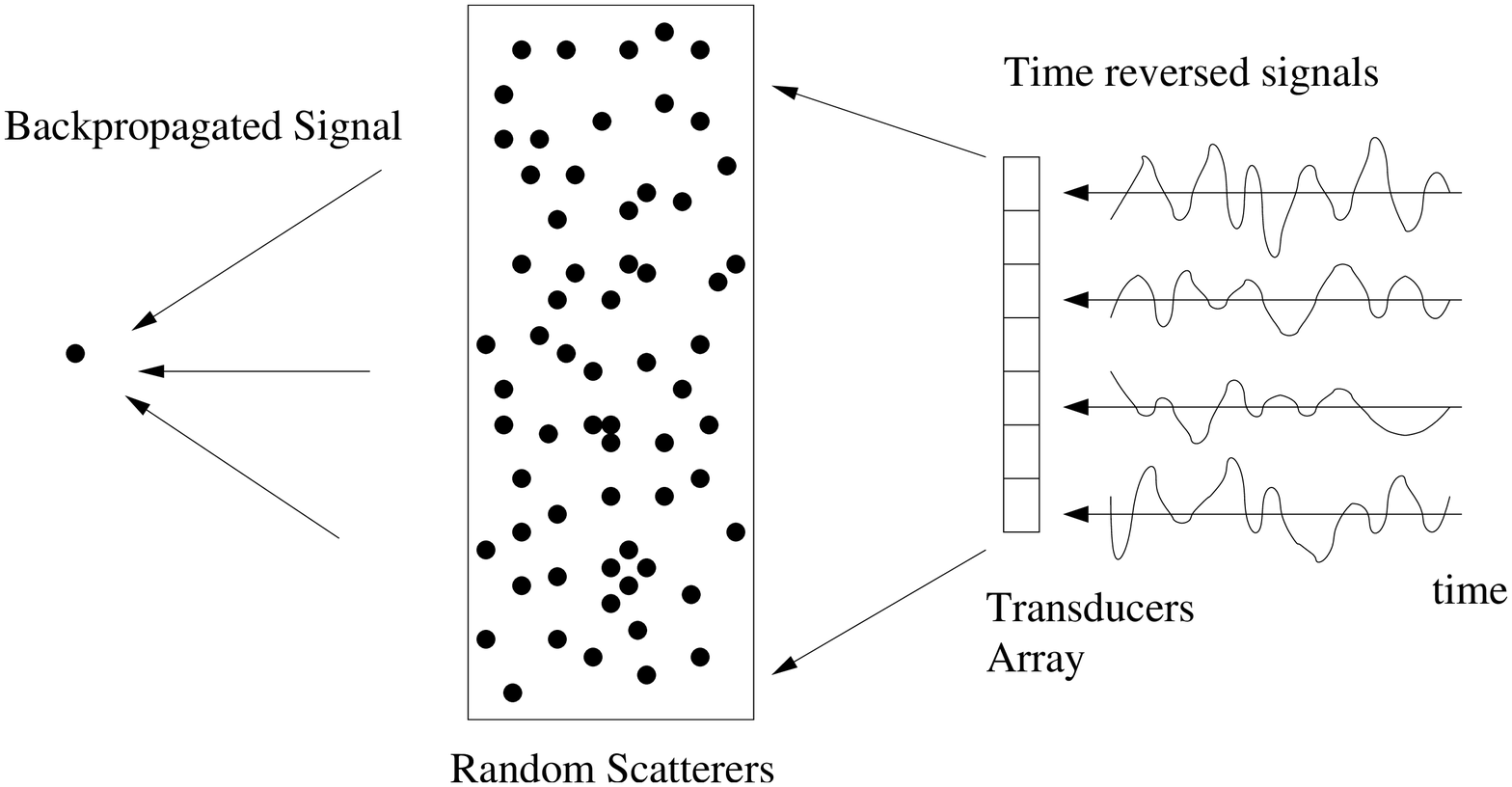}
    \caption{The Time Reversal Procedure. Top: Propagation of signal
      and measurements in time. Bottom: Time reversal of recorded
      signals and back-propagation into the medium.}
    \label{fig:exp}
  \end{center}
\end{figure}

A schematic description of the time
reversal procedure is depicted in Fig. \ref{fig:exp}.
Early experiments in time reversal acoustics are described in
\cite{DRF-PRL95}; see also the more recent papers
\cite{Fink-PT,Fink-01,Fink-Prada-01}. A very qualitative explanation
for the better refocusing observed in heterogeneous media is based on
{\em multipathing}. Since waves can scatter off a larger number of
heterogeneities, more paths coming from the source reach the recording
array, thus more is known about the source by the transducers than in
a homogeneous medium. The heterogeneous medium plays the role of a
lens that widens the aperture through which the array of receivers
sees the source.  Refocusing is also qualitatively justified by ray
theory (geometrical optics).  The phase shift caused by multiple
scattering is exactly compensated when the time reversed signal
follows the same path back to the source location. This phase
cancellation happens only at the source location.  The phase shift
along paths leading to other points in space is essentially random.
The interference of multiple paths will thus be constructive at the
source location and destructive anywhere else.  This explains why
refocusing at the source location is improved when the number of
scatterers is large.

As convincing as they are, the above explanations remain qualitative
and do not allow us to quantify how the refocused signal is modified
by the time reversal procedure.  Quantitative justifications require
to analyze wave propagation more carefully.  The first quantitative
description of time reversal was obtained in \cite{Fouque-Clouet} in
the framework of randomly layered media (see also the recent work
\cite{Ewodo}). That paper provides the first mathematical explanation
of two of the most prominent features of time reversal:
heterogeneities improve refocusing and refocusing occurs for almost
every realization of the random medium.  The first multi-dimensional
quantitative description of time reversal was obtained in
\cite{BPZ-JASA01} for the parabolic approximation, i.e., for waves
that propagate in a privileged direction with no backscattering (see also
\cite{PRS} for further analysis of time reversal in this regime). That
paper shows that the random medium indeed plays the role of a lens.
The back-propagated signal behaves as if the initial array were
replaced by another one with a much bigger effective aperture. In a
slightly different context, a recent paper \cite{BF-01} analyzes time
reversal in ergodic cavities.  There, wave mixing is created by
reflection at the boundary of a chaotic cavity, which plays a similar
role to the heterogeneities in a heterogeneous medium.

This paper generalizes the results of \cite{BPZ-JASA01} to the case of
general classical waves propagating in weakly fluctuating random
media. The main results are briefly summarized as follows. We first
show that refocusing in time reversal experiments may be understood in
the following three-step more general framework:
\begin{itemize}
\item[(i)]{A signal propagating from a localized source is recorded at
    a single time $T>0$ by an array of receivers.}
\item[(ii)]{The recorded signal is processed at the array location.}
\item[(iii)]{The processed signal is emitted from the array and
    propagates in the {\em same} medium during the same time $T$.}
\end{itemize}
The first main result is that the resulting signal will refocus at the
location of the original source for a large class of waves and a large
class of processings. The experiments described above correspond to
the specific processing of acoustic waves in which pressure is kept
unchanged and the direction of the acoustic field is reversed.

The second main result is a quantitative description of the
re-transmitted signal. We show that the re-propagated signal
$\bu^B(\vxi)$ at a point $\vxi$ near the source location can be
written in the high frequency limit as the following convolution of
the original source $\bS$
\begin{equation}
  \label{eq:refocintro}
\bu^B(\vxi)=({F}*{\bS})(\vxi).
\end{equation}
The kernel $F$ depends on the location of the recording array and on
the signal processing.  The quality of the refocusing depends on the
spatial decay of $F$.  It turns out that it can be
expressed in terms of the Wigner transform \cite{RPK-WM} of two
wave fields.  The decay properties of $F$ depend on the
smoothness of the Wigner transform in the phase space. The Wigner
transform in random media has been extensively studied
\cite{Erdos-Yau2,RPK-WM,Spohn}, especially in the high frequency
regime, when the wavelength of the initial signal is small compared to
the distance of propagation. It satisfies a radiative transport
equation, which is used to describe the evolution of the energy
density of waves in random media \cite{Ishimaru,RPK-WM,Sheng,Spohn}.
The transport equations possess a smoothing effect so that the Wigner
distribution becomes less singular in random media, which implies a
stronger decay of the convolution kernel $F$ and a better refocusing.
The diffusion approximation to the radiative transport equations
provides simple reconstruction formulas that can be used to quantify
the refocusing quality of the back-propagated signal.  This
construction applies to a large class of classical waves: acoustic,
electromagnetic, elastic, and others, and allows for a large class of
signal processings at the recording array.

Some results of this paper have been announced in \cite{BR-CRAS}.  The
concept of one-step time reversal emerged during early discussions
with Knut Solna. We also stress that the important property of
self-averaging of the time reversed signal (the refocused signal is
almost independent of the realization of the random medium) is not
analyzed in this paper. A formal explanation is given in
\cite{BPZ-JASA01,PRS} in the parabolic approximation.  Self-averaging
for classical waves will be addressed elsewhere.

This paper is organized as follows. Section \ref{sec:one-step} recalls
the classical setting of time reversal and introduces one-step time
reversal.  The re-transmitted signal and its relation to the Wigner
transform are analyzed in section \ref{sec:TRRM}.  A quantitative
description of acoustic wave refocusing in weakly fluctuating random
media is obtained by asymptotic analysis; see equations
(\ref{eq:filtersimp}) and (\ref{eq:filterdiff}) for an explicit
expression in the diffusion approximation. Section \ref{sec:ref}
generalizes the results in two ways. First, a more general signal
processing at the recording array is allowed, such as recording only
the pressure field of acoustic waves and not the velocity field.
Second, the re-transmission scheme is applied to more general waves
and the role of polarization and mode coupling is explained.

We would like to thank Knut Solna for fruitful discussions during the
preparation of this work. We are indebted to George Papanicolaou for
his contributions to the analysis of time reversal, which lie at the
core of this paper.  This work would also not have been possible without
the numerous exchanges we benefited from at the Stanford MGSS summer
school.
 
\section{Classical Time Reversal and One-Step Time Reversal}
\label{sec:one-step}

Propagation of acoustic waves is described by a system of equations
for the pressure $p(t,\vx)$ and acoustic velocity $\bv(t,\vx)$:
\begin{eqnarray}\label{eq:acoust}
&&\rho(\vx)\pdr{\bv}{t}+\nabla p=0\\
&&\kappa(\vx)\pdr{p}{t}+\nabla\cdot\bv=0,\nonumber
\end{eqnarray}
with suitable initial conditions and where $\rho(\bx)$ and
$\kappa(\bx)$ are density and compressibility of the underlying
medium, respectively.  These equations can be recast as
the following linear hyperbolic system
\begin{equation}\label{eq:general}
A(\vx)\pdr{\vu}{t}+D^j\pdr{\vu}{x^j}=0,~~~\vx\in{\mathbb R}^3
\end{equation}
with the vector $\vu=(\bv,p)\in{\mathbb C}^4$. The matrix
$A=\hbox{Diag}(\rho,\rho,\rho,\kappa)$ is positive definite. The
$4\times 4$ matrices $D^j$, $j=1,2,3$, are symmetric and given by
$D_{mn}^j=\delta_{m4}\delta_{nj}+ \delta_{n4}\delta_{mj}$. We use the
Einstein convention of summation over repeated indices.

The time reversal experiments in \cite{DRF-PRL95} consist of two
steps. First, the direct problem
\begin{eqnarray}\label{eq:direct}
&&\!\!A(\vx)\pdr{\vu}{t}+D^j\pdr{\vu}{x^j}=0,~0\le t\le T\\
&&\!\!\vu(0,\vx)=\bS(\vx)\nonumber
\end{eqnarray}
with a localized source $\bS$ centered at a point $\vx_0$ is solved.
The signal is recorded during the period of time $0\le t\le T$ by an
array of receivers located at $\Omega\subset\Rm^3$. Second, the signal
is time reversed and re-emitted into the medium. Time reversal is
described by multiplying $\bu=(\bv,p)$ by the matrix
$\Gamma=\hbox{Diag}(-1,-1,-1,1)$.  The back-propagated signal solves
\begin{eqnarray}\label{eq:reverse-pr}
&&\pdr{\vu}{t}+A^{-1}(\vx)D^j\pdr{\vu}{x^j}=\frac{1}{T}{\bf R}(2T-t,\vx),
~T\le t\le 2T\\
&&\vu(T,\vx)=0\nonumber
\end{eqnarray}
with the source term
\begin{equation}\label{eq:rev-source}
{\bf R}(t,\vx)=\Gamma{\vu(t,\vx)}\chi(\vx).
\end{equation}
The function $\chi(\vx)$ is either the characteristic function of the
set where the recording array is located, or some other function that
allows for possibly space-dependent amplification of the
re-transmitted signal.

The back-propagated signal is then given by $\vu(2T,\vx)$.  We can
decompose it as
\begin{eqnarray}\label{duhamel}
&&\vu(2T,\vx)
=\frac 1T\int_0^T\!\!\!ds~ \bw(s,\vx;s),
\end{eqnarray}
where the vector-valued function $\bw(t,\vx;s)$ solves the initial
value problem
\begin{eqnarray*}
&&A(\bx)\pdr{\bw(t,\vx;s)}{t}+D^j\pdr{\bw(t,\vx;s)}{x^j}=0,~~0\le t\le s\\
&&\bw(0,\vx;s)={\bf R}(s,\vx).
\end{eqnarray*}
We deduce from (\ref{duhamel}) that it is sufficient to analyze the
refocusing properties of $\bw(s,\vx;s)$ for $0\leq s \leq T$ to obtain
those of $\vu(2T,\vx)$.  For a fixed value of $s$, we call the
construction of $\bw(s,\vx;s)$ one-step time reversal.

We define one-step time reversal more generally as follows.  The
direct problem (\ref{eq:direct}) is solved until time $t=T$ to yield
$\bu(T^-,\vx)$. At time $T$, the signal is recorded and processed.
The processing is modeled by an amplification function $\chi(\bx)$, a
blurring kernel $f(\bx)$, and a (possibly spatially varying) time
reversal matrix $\Gamma$.  After processing, we have
\begin{equation}
  \label{eq:buT+}
  \bu(T^+,\bx)={\Gamma}(f*(\chi\bu))(T^-,\bx)\chi(\bx).
\end{equation}
The processed signal then propagates during the same time $T$:
\begin{eqnarray}\label{reverse-onestep}
&&\!\!A(\vx)\pdr{\vu}{t}+D^j\pdr{\vu}{x^j}=0,~T\le t\le 2T\\
&&\!\!\bu(T^+,\vx)={\Gamma}(f*(\chi\bu))(T^-,\vx)\chi(\vx).\nonumber
\end{eqnarray}
The main question is whether $\bu(2T,\vx)$ refocuses at the location
of the original source $\bS(\vx)$ and how the original signal has been
modified by the time reversal procedure. Notice that in the case of
full ($\Omega=\Rm^3$) and exact ($f(\bx)=\delta(\bx)$) measurements
with $\Gamma=\mbox{Diag}(-1,-1,-1,1)$, the time-reversibility of
first-order hyperbolic systems implies that $\bu(2T,\bx)=\Gamma
\bS(\bx)$, which corresponds to exact refocusing.
When only partial measurements are available we shall see in the following
sections that $\vu(2T,\vx)$ is closer to $\Gamma\bS(\bx)$ when propagation
occurs in a heterogeneous medium than in a homogeneous medium.

\begin{figure}[htbp]
  \begin{center}
   \hspace{-1cm}\includegraphics[height=4.5cm]{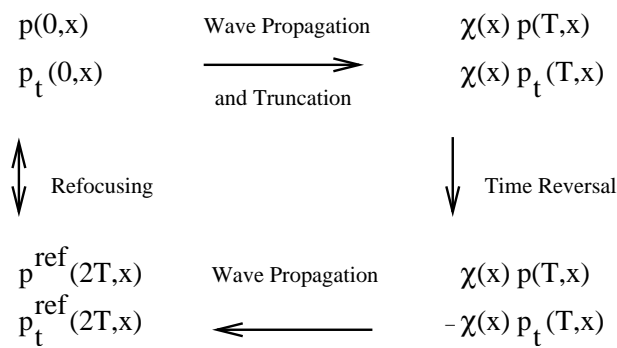}
    \caption{The One-Step Time Reversal Procedure. Here,
   $p_t$ denotes $\pdr{p}{t}$.}
    \label{fig:onestep}
  \end{center}
\end{figure}

\begin{figure}[htbp]
  \centering
  $\mbox{}\!\!\!\!\!\!\!\!\!\!\!\!\!\!\!\!\!\!\!\!\!\!\!\!\!\!\!$
  \includegraphics[height=5.6cm]{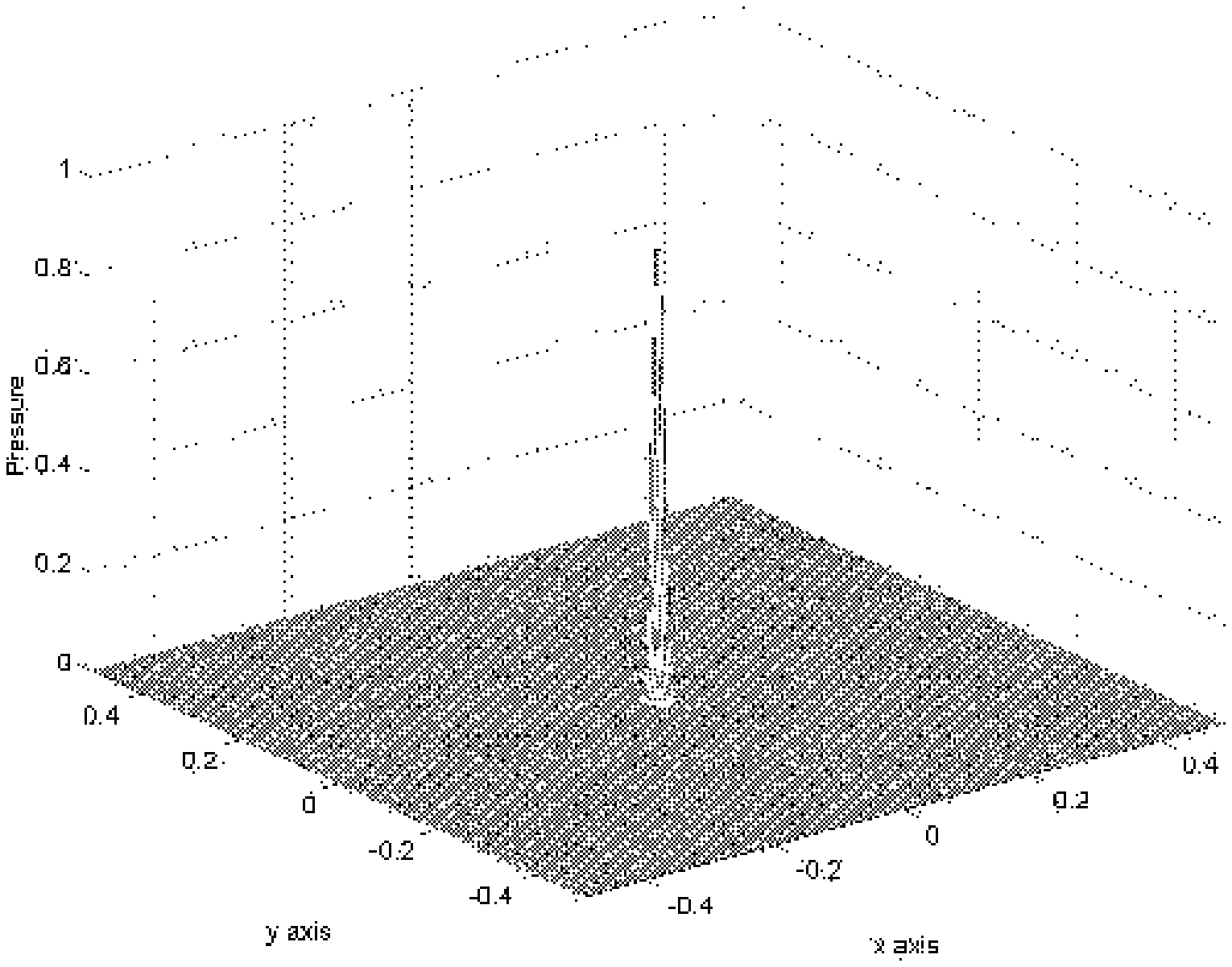}$\!\!\!\!\!\!\!\!\!\!\!\!$
  \includegraphics[height=5.6cm]{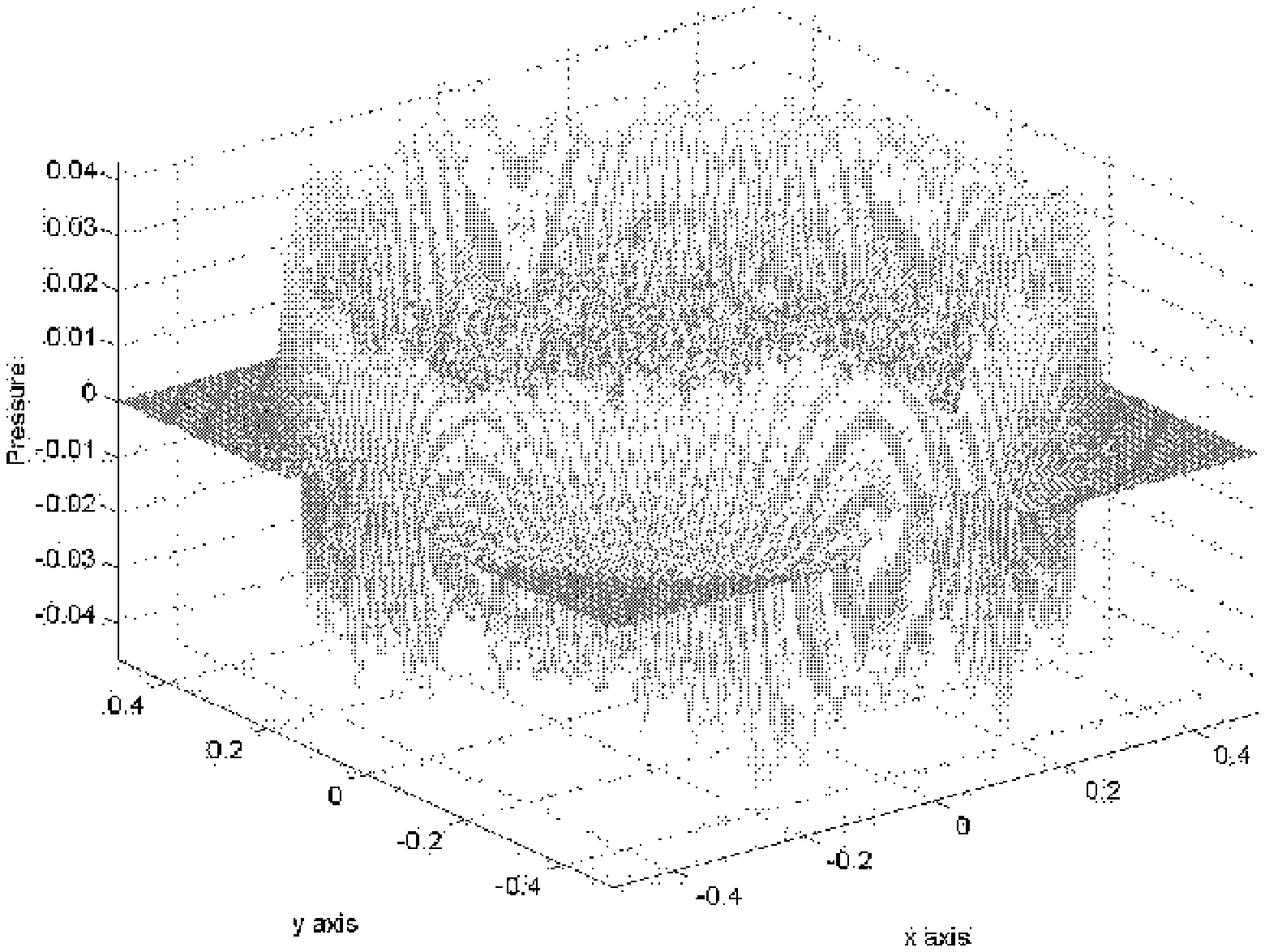}
    $\mbox{}\!\!\!\!\!\!\!\!\!\!\!\!\!\!\!\!\!\!\!\!\!\!\!\!\!\!\!$\\
  $\mbox{}\!\!\!\!\!\!\!\!\!\!\!\!\!\!\!\!\!\!\!\!\!\!\!\!\!\!\!$
  \includegraphics[height=5.6cm]{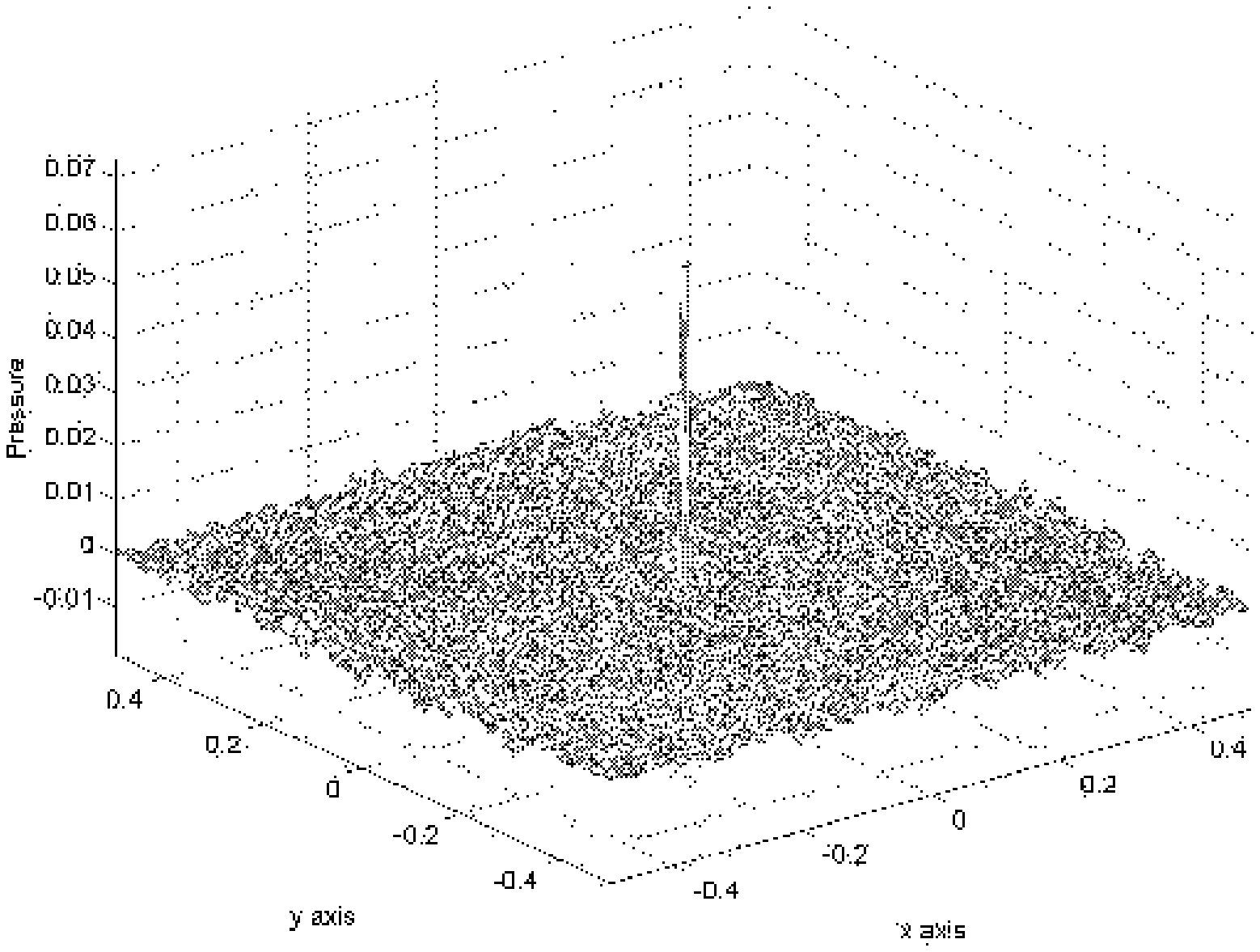}$\!\!\!\!\!\!\!\!\!\!\!\!$
  \includegraphics[height=5.6cm]{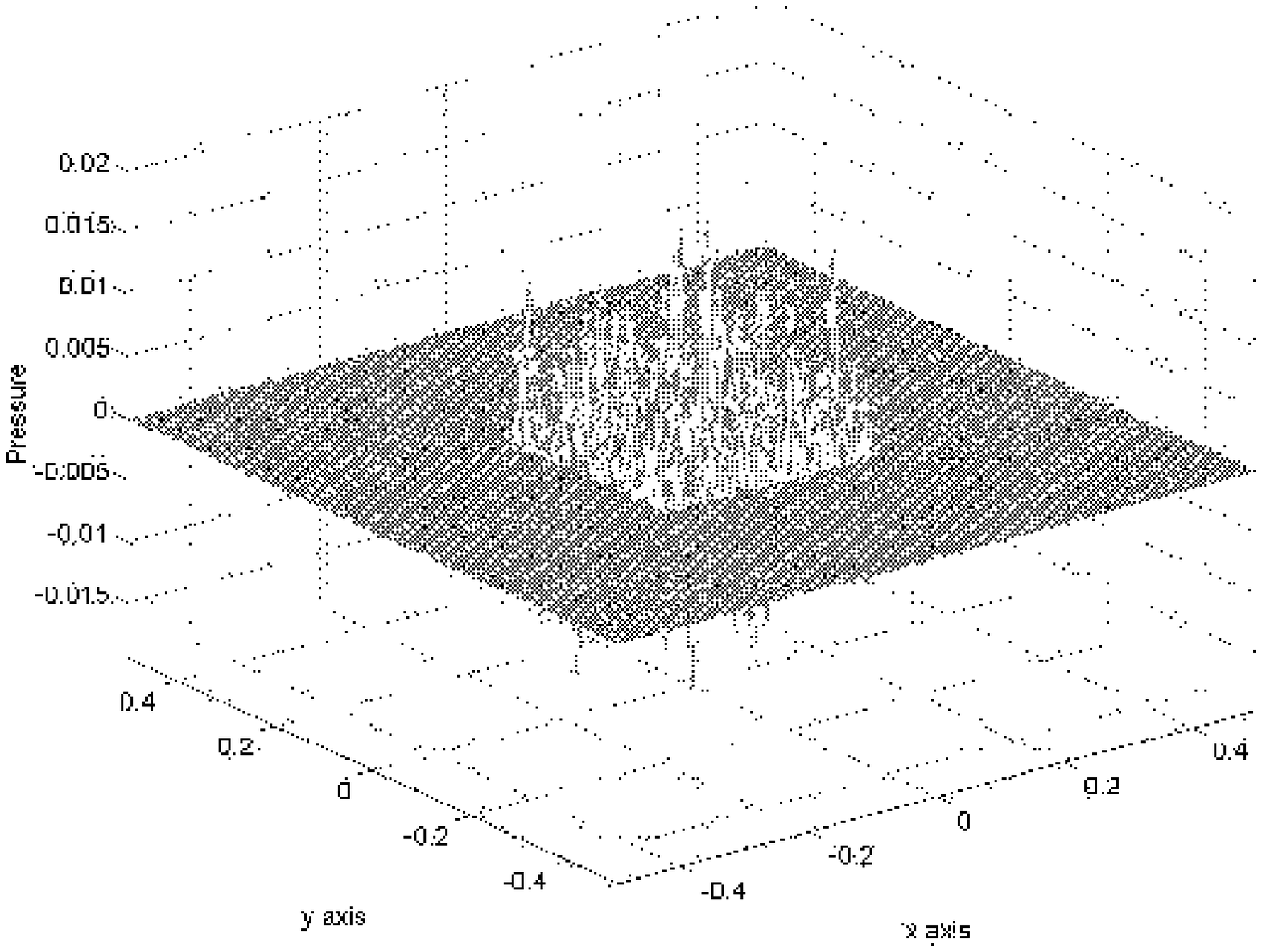}
    $\mbox{}\!\!\!\!\!\!\!\!\!\!\!\!\!\!\!\!\!\!\!\!\!\!\!\!\!\!\!$
  \caption{Numerical experiment of the one-step time reversal procedure.
    Top Left: initial condition $p(0,\bx)$, a peaked Gaussian of
    maximal amplitude equal to $1$. Top Right: forward solution
    $p(T^-,\bx)$, of maximal amplitude $0.04$. Bottom Right: recorded
    solution $p(T^+,\bx)$, of maximal amplitude $0.015$ on the domain
    $\Omega=(-1/6,1/6)^2$. Bottom Left: back-propagated solution
    $p(2T,\bx)$, of maximal amplitude $0.07$. }
  \label{fig:num1}
\end{figure}

The pressure field $p(t,\vx)$ satisfies
the following scalar wave equation
\begin{equation}
  \label{eq:wave}
\frac{\partial^2p}{\partial t^2}-
\frac{1}{\kappa(\bx)}\nabla\cdot\left(\frac{1}{\rho(\bx)}\nabla p\right)=0.
\end{equation}
A schematic description of the one-step procedure for the wave
equation is presented in Fig.  \ref{fig:onestep}.  This is the
equation solved in the numerical experiments presented in this paper.
The details of the numerical setting are described in the appendix.
A numerical experiment for the one-step time reversal procedure is
shown in Fig. \ref{fig:num1}.
In the numerical simulations, there is no blurring,
$f(\bx)=\delta(\bx)$, and the array of receivers is the domain
$\Omega=(-1/6,1/6)^2$ ($\chi(\bx)$ is the characteristic function of
$\Omega$).  Note that the truncated signal does not retain any
information about the ballistic part (the part that propagates without
scattering with the underlying medium).  In homogeneous medium, the
truncated signal would then be identically zero and no refocusing
would be observed.  The interesting aspect of time reversal is that
a coherent signal emerges at time $2T$ out of a signal at time $T^+$
that seems to have no useful information.

\section{Theory of Time Reversal in Random Media}
\label{sec:TRRM}

Our objective is now to present a theory that explains in a
quantitative manner the refocusing properties described in the
preceding sections. We consider here the one-step time reversal for
acoustic wave. Generalizations to other types of waves and more
general processings in (\ref{reverse-onestep}) are given in section
\ref{sec:ref}.

\subsection{Refocused Signal}
\label{sec:ub}

We recall that the one-step time reversal procedure consists of
letting an initial pulse $\bS(\bx)$ propagate according to
(\ref{eq:direct}) until time $T$,
\begin{displaymath}
  \bu(T^-,\bx)=\dint_{\Rm^\d} G(T,\bx;\bz)\bS(\bz) d\bz,
\end{displaymath}
where $G(T,\vx;\vz)$ is the Green's matrix solution of
\begin{eqnarray}\label{green's}
&&\!\!\!
A(\vx)\pdr{G(t,\vx;\vy)}{t}+D^j\pdr{G(t,\vx;\vy)}{x^j}=0,~0\le t\le T\\
&&\!\!\! G(0,\vx;\vy)=I\delta(\vx-\vy).\nonumber
\end{eqnarray}
At time $T$, the ``intelligent'' array reverses the signal. For
acoustic pulses, this means keeping pressure unchanged and reversing
the sign of the velocity field. The array of receivers is located in
$\Omega\subset\Rm^\d$.  The amplification function $\chi(\bx)$ is an
arbitrary bounded function supported in $\Omega$, such as its
characteristic function ($\chi(\bx)=1$ for $\bx\in\Omega$ and
$\chi(\bx)=0$ otherwise) when all transducers have the same
amplification factor.  We also allow for some blurring of the recorded
data modeled by a convolution with a function $f(\bx)$. The case
$f(\bx)=\delta(\bx)$ corresponds to exact measurements. Finally, the
signal is time reversed, that is, the direction of the acoustic
velocity is reversed. Here, the operator $\Gamma$ in (\ref{eq:buT+})
is simply multiplication by the matrix
\begin{equation}
  \label{eq:gamma}
  \Gamma=\mbox{Diag}(-1,-1,-1,1).
\end{equation}
The signal at time $T^+$ after time reversal takes then the form
\begin{equation}
  \label{eq:prop}
  \bu(T^+,\bx)=
    \dint_{\Rm^{6}} \Gamma G(T,\by';\bz)
      \chi(\bx)\chi(\by') f(\bx-\by') \bS(\bz) d\bz d\by'.
\end{equation}

The last step (\ref{reverse-onestep}) consists of letting the time
reversed field propagate through the random medium until time $2T$. To
compare this signal with the initial pulse $\bS$, we need to reverse
the acoustic velocity once again, and define
\begin{equation}
  \label{eq:refoc}
\bu^B(\vx)=    \Gamma\bu(2T,\bx)=\dint_{\Rm^{9}} \Gamma G(T,\bx;\by)
       \Gamma G(T,\by';\bz)
      \chi(\by)\chi(\by') f(\by-\by') \bS(\bz) d\by d\by'd\bz.
\end{equation}

The time reversibility of first-order hyperbolic systems implies that
$\bu^B(\bx)=\bS(\bx)$ when $\Omega=\Rm^\d$, $\chi\equiv 1$, and
$f(\bx)=\delta(\bx)$, that is, when full and non-distorted
measurements are available.  It remains to understand which features
of $\bS$ are retained by $\bu^B(\bx)$ when only partial measurement is
available.

\subsection{Localized Source and Scaling}
\label{sec:loc-source}

We consider an asymptotic solution of the time reversal problem
(\ref{eq:direct}), (\ref{reverse-onestep}) when the support $\lambda$
of the initial pulse $\bS(\vx)$ is much smaller than the distance $L$
of propagation between the source and the recording array:
$\eps=\lambda/L\ll 1$. We also take the size $a$ of the array
comparable to $L$: $a/L=O(1)$.  We assume that the time $T$ between
the emission of the original signal and recording is of order $L/c_0$,
where $c_0$ is a typical speed of propagation of the acoustic wave.
We consequently consider the initial pulse to be of the form
\begin{displaymath}
  \vu(0,\vx)=\bS(\dfrac{\bx-\bx_0}{\eps})
\end{displaymath}
in non-dimensionalized variables $\vx'=\vx/L$ and $t'=t/(L/c_0)$.
We drop primes to simplify notation.  Here $\vx_0$ is the
location of the source.  The transducers obviously have to be capable
of capturing signals of frequency $\eps^{-1}$ and blurring should
happen on the scale of the source, so we replace $f(\bx)$ by
$\eps^{-\d}f(\eps^{-1}\bx)$. Finally, we are interested in the
refocusing properties of $\bu^B(\vx)$ in the vicinity of $\bx_0$. We
therefore introduce the scaling $\bx=\bx_0+\eps\Bxi$. With these
changes of variables, expression (\ref{eq:refoc}) is recast as
 \begin{eqnarray}
   \label{eq:refoceps}
    &&\bu^B(\Bxi;\bx_0)=\Gamma\bu(2T,\bx_0+\eps\vxi)\\
 &&~~~~~~~~~~~~=\dint_{\Rm^{9}}
     \Gamma G(T,\bx_0+\eps\Bxi;\by)
        \Gamma G(T,\by';\bx_0+\eps \bz)
     \chi(\by,\by')  \bS(\bz)
     d\by d\by' d\bz,\nonumber
 \end{eqnarray}
 where
 \begin{equation}\label{chieps}
 \chi(\by,\by')= \chi(\by)\chi(\by')f(\dfrac{\by-\by'}{\eps}).
 \end{equation}
In the sequel we will also allow the medium to vary on a scale
comparable to the source scale $\eps$. Thus the Green's function $G$ and
the matrix $A$ depend on $\eps$. We do not make this dependence
explicit to simplify notation.  We are interested in the limit of
$\bu^B(\Bxi;\bx_0)$ as $\eps\to0$.  The scaling considered here is
well adapted both to the physical experiments in \cite{DRF-PRL95}
and the numerical experiments in Fig. \ref{fig:num1}.

\subsection{Adjoint Green's Function}
\label{sec:adjoint}

The analysis of the re-propagated signal relies on the study of the
two point correlation at nearby points of the Green's matrix in
(\ref{eq:refoceps}).  There are two undesirable features in
(\ref{eq:refoceps}). First, the two nearby points $\bx_0+\eps\Bxi$ and
$\bx_0+\eps\bz$ are terminal and initial points in their respective
Green's matrices. Second, one would like the matrix $\Gamma$ between the
two Green's matrices to be outside of their product.  However, $\Gamma$ and
$G$ do not commute. For these reasons, we introduce the {\em adjoint}
Green's matrix, solution of
\begin{equation}
  \label{eq:adjointGreen}
  \begin{array}{l}
  \pdr{G_*(t,\bx;\by)}{t} A(\bx) + \pdr{G_*(t,\bx;\by)}{x^j}D^j =0 \\
  G_*(0,\bx;\by)=A^{-1}(\bx)\delta(\bx-\by).
  \end{array}
\end{equation}
We now prove that
\begin{equation}
  \label{eq:relgreen}
  G_*(t,\bx;\by)=\Gamma G(t,\by;\bx)A^{-1}(\bx) \Gamma.
\end{equation}
Note that for all initial data
$\bS(\vx)$, the solution $\bu(t,\bx)$ of (\ref{eq:direct})
satisfies
\begin{displaymath}
   \bu(t,\bx) =\dint_{\Rm^\d} G(t-s,\bx;\by) \bu(s,\by) d\by
\end{displaymath}
for all $0\leq s\leq t\leq T$ since the coefficients in (\ref{eq:direct})
are time-independent. Differentiating the above with respect to $s$
and using (\ref{eq:direct}) yields
\begin{displaymath}
  0=\dint_{\Rm^\d} \Big(-\pdr{G(t-s,\bx;\by)}{t}\bu(s,\by)
    - G(t-s,\bx;\by)A^{-1}(\by)D^j \pdr{\bu(s,\by)}{y^j}\Big)d\by
\end{displaymath}
Upon integrating by parts and letting $s=0$, we get
\begin{displaymath}
  0=\dint_{\Rm^\d} \Big(-\pdr{G(t,\bx;\by)}{t} +
  \pdr{}{y^j}\left[G(t,\bx;\by)A^{-1}(\vy)D^j\right]\Big)\bS(\by)d\by.
\end{displaymath}
Since the above relation holds for all test functions $\bS(\by)$,
we deduce that
\begin{equation}
  \label{eq:Giny}
  \pdr{G(t,\bx;\by)}{t} - \pdr{}{y^j}\left[G(t,\bx;\by)A^{-1}(\by)D^j\right]=0.
\end{equation}
Interchanging $\bx$ and $\by$ in the above equation and multiplying it
on the left and the right by $\Gamma$, we obtain that
\begin{equation}
  \label{eq:GtoG*}
  \pdr{}{t}\left[\Gamma G(t,\by;\bx) A^{-1}(\bx)\right]A(\bx)\Gamma
 -\pdr{}{x^j}\left[\Gamma G(t,\by;\bx) A^{-1}(\bx)\right] D^j\Gamma=0.
\end{equation}
We remark that
\begin{equation}
  \label{eq:commutgamma}
  \Gamma D^j=-D^j\Gamma\qquad \mbox{ and } \qquad
  \Gamma A(\bx)=A(\bx)\Gamma,
\end{equation}
so that
\begin{displaymath}
 \pdr{}{t}\left[\Gamma G(t,\by;\bx) A^{-1}(\bx)\Gamma\right]A(\bx)
 +\pdr{}{x^j}\left[\Gamma G(t,\by;\bx) A^{-1}(\bx)\Gamma\right] D^j=0
\end{displaymath}
with $\Gamma G(0,\by;\bx)
A^{-1}(\bx)\Gamma=A^{-1}(\vx)\delta(\vx-\vy)$.  Thus
(\ref{eq:relgreen}) follows from the uniqueness of the solution to the
above hyperbolic system with given initial conditions.  We can now
recast (\ref{eq:refoceps}) as
\begin{eqnarray}
  \label{eq:refoceps2}
   \begin{array}{ll}
    \bu^B(\Bxi;\bx_0) \,=\, \dint_{\Rm^{9}} &\!\!\!
        \Gamma G(T,\bx_0+\eps\Bxi;\by)
       G_*(T,\bx_0+\eps \bz;\by') \Gamma \\
   &\!\!\! \times\chi(\by)\chi(\by')
    f(\dfrac{\by-\by'}{\eps}) A(\bx_0+\eps\bz)
    \bS(\bz) d\by d\by' d\bz.
   \end{array}
\end{eqnarray}

One further simplifies (\ref{eq:refoceps2}) with the help of the
auxiliary matrix-valued functions $Q(t,\vx;\bq)$ and $Q_*(t,\vx,\bq)$
defined by
\begin{equation}
  \label{eq:QandQ*}
  \begin{array}{rcl}
  Q(T,\bx;\bq)&=&\dint_{\Rm^\d}
    G(T,\bx;\by) \chi(\by)e^{i\bq\cdot\by/\eps} d\by, \\
  Q_*(T,\bx;\bq)&=&\dint_{\Rm^\d}G_*(T,\bx;\by)
     \chi(\by)e^{-i\bq\cdot\by/\eps} d\by.
  \end{array}
\end{equation}
They solve the hyperbolic equations (\ref{eq:direct}) and
(\ref{eq:adjointGreen}) with initial conditions given by
$Q(0,\vx;\bq)=\chi(\bx)e^{i\bq\cdot\bx/\eps}I$ and
$Q_*(0,\vx;\bq)=A^{-1}(\bx)\chi(\bx)e^{-i\bq\cdot\bx/\eps}$,
respectively.  Thus (\ref{eq:refoceps2}) becomes
\begin{equation}
  \label{eq:refoceps3}
   \bu^B(\Bxi;\bx_0)\!\!=\!\!\dint_{\Rm^{6}}
    \!\!
    \Gamma Q(T,\bx_0+\eps\Bxi;\bq)
       Q_*(T,\bx_0+\eps \bz;\bq) \Gamma A(\vx_0+\eps\vz)
    \bS(\bz) \hat f(\bq) \frac{d\bq d\bz}{(2\pi)^\d},
\end{equation}
where $\hat f(\bq)=\int_{\Rm^\d}e^{-i\bq\cdot\bx}f(\bx)d\bx$ is the Fourier
transform of $f(\bx)$.

\subsection{Wigner Transform}
\label{sec:wigner}

The back-propagated signal in (\ref{eq:refoceps3}) now has the
suitable form to be analyzed in the Wigner transform formalism
\cite{GMMP,RPK-WM}. We define
\begin{equation}
  \label{eq:Wigner}
  W_\eps(t,\bx,\bk)=\dint_{\Rm^\d} \hat f(\bq) U_\eps(t,\bx,\bk;\bq) {d\bq},
\end{equation}
where
\begin{equation}
  \label{eq:WignerU}
  U_\eps(t,\bx,\bk;\bq)=
   \dint_{\Rm^\d}e^{i\bk\cdot\by} Q(t,\bx-\dfrac{\eps\by}{2};\bq)
  Q_*(t,\bx+\dfrac{\eps\by}{2};\bq) \dfrac{d\by}{(2\pi)^\d}.
\end{equation}
Taking inverse Fourier transform we verify that
\begin{displaymath}
  Q(t,\bx;\bq)Q_*(t,\by;\bq)=
\dint_{\Rm^\d} e^{-i\bk\cdot(\by-\bx)/\eps}
U_\eps(t,\dfrac{\bx+\by}{2},\bk;\bq) d\bk,
\end{displaymath}
hence
\begin{equation}
  \label{eq:refoceps4}
   \bu^B(\Bxi;\bx_0)=\dint_{\Rm^{6}}e^{i\bk\cdot(\Bxi-\bz)}
\Gamma W_\eps(T,\bx_0+\eps\dfrac{\bz+\Bxi}{2},\bk) \Gamma A(\bx_0+\eps\bz)
      \bS(\bz) \frac{d\bz d\bk}{(2\pi)^\d}.
\end{equation}

We have thus reduced the analysis of $\bu(\Bxi;\bx_0)$ as $\eps\to0$
to that of the asymptotic properties of the Wigner transform $W_\eps$.
The Wigner transform has been used extensively in the study of wave
propagation in random media, especially in the derivation of radiative
transport equations modeling the propagation of high frequency waves.
We refer to \cite{GMMP,LP,RPK-WM}. Note that in the usual definition
of the Wigner transform, one has the adjoint matrix $Q^*$ in place of
$Q_*$ in (\ref{eq:WignerU}). This difference is not essential since
$Q_*$ and $Q^*$ satisfy the same evolution equation, though with
different initial data.

The main reason for using the Wigner transform in (\ref{eq:refoceps4})
is that $W_\eps$ has a weak limit $W$ as $\eps\to 0$.  Its existence
follows from simple a priori bounds for $W_\eps(t,\bx,\bk)$. Let us
introduce the space ${\cal A}$ of matrix-valued functions
$\phi(\vx,\bk)$ bounded in the norm $\|\cdot\|_{{\cal A}}$ defined by
\begin{displaymath}
  \|\phi\|_{{\cal A}} = \dint_{\Rm^{\d}} \sup_{\vx} \|\tilde\phi(\bx,\by)\|
   d\by ,\qquad \mbox{ where } \quad
  \tilde\phi(\bx,\by)=\dint_{\Rm^\d} e^{-i\bk\cdot \by} \phi(\bx,\bk) d\bk.
\end{displaymath}
We denote by ${\cal A}'$ its dual space, which is a space of
distributions large enough to contain matrix-valued bounded measures,
for instance.  We then have the following result:
\begin{lemma}
  \label{lem:bound} Let $\chi(\bx)\in L^2(\Rm^\d)$ and $\hat f(\bq)\in
  L^1(\Rm^\d)$. Then there is a constant $C>0$ independent of
  $\eps>0$ and $t\in[0,\infty)$ such that for all $t\in[0,\infty)$, we
  have $\|W_\eps(t,\bx,\bk)\|_{{\cal A}'}<C$.
\end{lemma}\\
The proof of this lemma is essentially contained in \cite{GMMP,LP};
see also \cite{BR-CRAS}. One may actually get $L^2$-bounds for
$W_\eps$ in our setting because of the regularizing effect of $\hat f$
in (\ref{eq:Wigner}) but this is not essential for the purposes of
this paper. We therefore obtain the existence of a subsequence
$\eps_k\to0$ such that $W_{\eps_k}$ converges weakly to a distribution
$W\in{\cal A}'$. Moreover, an easy calculation shows that at time
$t=0$, we have
\begin{equation}\label{in-w}
  W(0,\bx_0,\bk)=|\chi(\bx_0)|^2 A_0^{-1}(\bx_0) \hat f(\bk).
\end{equation}
Here, $A_0=A$ when $A$ is independent of $\eps$, and
$A_0=\lim\limits_{\eps\to0} A_\eps$ if we assume that the family of
matrices $A_\eps(\bx)$ is uniformly bounded and continuous with the
limit $A_0$ in ${\cal C}({\mathbb R}^d)$. These assumptions on $A_\eps$ are
sufficient to deal with the radiative transport regime we will
consider in section \ref{sec:rte}.  Under the same assumptions on
$A_\eps$, we have the following result.
\begin{proposition}
  \label{ref:prop}
  The back-propagated signal $\bu^B(\Bxi;\bx_0)$ given by
  (\ref{eq:refoceps4}) converges weakly in ${\cal S}'(\Rm^\d\times
  \Rm^\d)$ as $\eps\to 0$ to the limit
  \begin{equation}
    \label{eq:limitub}
    \bu^B(\Bxi;\bx_0)=\dint_{\Rm^{6}} e^{i\bk\cdot(\Bxi-\bz)}
 \Gamma W(T,\bx_0,\bk) \Gamma A_0(\bx_0) \bS(\bz) \frac{d\bz d\bk}{(2\pi)^\d}.
  \end{equation}
\end{proposition}
The proof of this proposition is based on taking the duality product
of $\bu^B(\Bxi;\bx_0)$ with a vector-valued test function
$\bphi(\Bxi;\bx_0)$ in ${\cal S}(\Rm^\d\times\Rm^\d)$. After a change of
variables we obtain
\begin{math}
  \langle \bu^B,\bphi\rangle = \langle W_\eps,Z_\eps\rangle.
\end{math}
Here the duality product for matrices is given by the trace $\langle
A,B\rangle=\sum_{i,k} \langle A_{ik},B_{ik}\rangle$, and
\begin{equation}
  \label{eq:Zeps}
  Z_\eps(\bx_0, \bk)=\dint_{\Rm^{6}} e^{i\bk\cdot(\bz-\Bxi)}
   \Gamma \bphi(\Bxi,\bx_0-\eps\dfrac{\bz+\Bxi}{2})\bS^*(\bz)
A_\eps(\bx_0+\eps\dfrac{\bz-\Bxi}{2})\Gamma \frac{d\bz d\Bxi}{(2\pi)^\d}.
\end{equation}
Defining $Z$ as the limit of $Z_\eps$ as $\eps\to0$ by replacing
formally $\eps$ by $0$ in the above expression, (\ref{eq:limitub})
follows from showing that $\|Z_\eps-Z\|_{{\cal A}}\to0$ as $\eps\to0$.
This is straightforward and we omit the details.

The above proposition tells us how to reconstruct the back-propagated
solution in the high frequency limit from the limit Wigner matrix $W$.
Notice that we have made almost no assumptions on the medium described
by the matrix $A_\eps(\bx)$. At this level, the medium can be either
homogeneous or heterogeneous. Without any further assumptions, we can
also obtain some information about the matrix $W$.  Let us define the
dispersion matrix for the system (\ref{eq:direct}) as \cite{RPK-WM}
\begin{equation}
  \label{eq:L}
  L(\bx,\bk)=A_0^{-1}(\bx) k_j D^j.
\end{equation}
It is given explicitly by
\[
L(\bx,\bk)=\left(\begin{matrix} 0 & 0 & 0 & k_1/\rho(\bx)\cr
                       0 & 0 & 0 & k_2/\rho(\bx)\cr
                       0 & 0 & 0 & k_3/\rho(\bx)\cr
                       k_1/\kappa(\bx) & k_2/\kappa(\bx) & k_3/\kappa(\bx)
    & 0\cr\end{matrix}
\right).
\]
The matrix $L$ has a double eigenvalue $\omega_0=0$ and two simple
eigenvalues $\omega_{\pm}(\bx,\bk)=\pm c(\bx)|\bk|$, where
$c(\bx)=1/\sqrt{\rho(\bx)\kappa(\bx)}$ is the speed of sound. The
eigenvalues $\omega_{\pm}$ are associated with eigenvectors
$\bb_{\pm}(\bx,\bk)$ and the eigenvalue $\omega_0=0$ is associated
with the eigenvectors $\bb_j(\bx,\bk)$, $j=1,2$. They are given by
\begin{equation}
  \label{eq:bbpm} 
   \bb_{\pm}(\bx,\bk)=\left(
    \begin{array}{c}
       \pm\dfrac{\hat \bk}{\sqrt{2\rho(\bx)}}  \\
        \dfrac{1}{\sqrt{2\kappa(\bx)}}
    \end{array} \right), \qquad \qquad
    \bb_j(\bx,\bk)=\left(
    \begin{array}{c}
       \dfrac{{\bf z}^j(\bk)}{\sqrt{\rho(\bx)}}  \\ 0
    \end{array} \right),
\end{equation}
where $\hat\bk=\bk/|\bk|$ and $\vz^{1}(\bk)$ and $\vz^2(\bk)$ are
chosen so that the triple $(\vhatk,\vz^1(\bk),\vz^2(\bk))$ forms an
orthonormal basis.  The eigenvectors are normalized so that
\begin{equation}
  \label{eq:scalarA0}
(A_0(\vx) \bb_j(\bx,\bk) \cdot \bb_k(\bx,\bk)) = \delta_{jk},
\end{equation}
for all $j,k\in J=\{+,-,1,2\}$.  The space of $4\times4$ matrices is
clearly spanned by the basis $\bb_{j}\otimes\bb_{k}$.  We then have
the following result:
\begin{proposition}\label{prop:decomp}
There exist scalar distributions $a_\pm$ and $a_{0}^{mn}$, $m,n=1,2$
so that the limit Wigner distribution matrix can be decomposed as
\begin{eqnarray}\label{eq:decomp}
&& W(t,\bx,\bk)=\dsum_{j,m=1}^2 a_{0}^{jm}(t,\bx,\bk)
\bb_j(\bx,\bk)\otimes\bb_m(\bx,\bk)\\
&&\quad +\,a_+(t,\bx,\bk)\bb_+(\bx,\bk)\otimes\bb_+(\bx,\bk)
+a_-(t,\bx,\bk)\bb_-(\bx,\bk)\otimes\bb_-(\bx,\bk).\nonumber
\end{eqnarray}
\end{proposition}
The main result of this proposition is that the cross terms
$\bb_j\otimes\bb_k$ with $\omega_j\not=\omega_k$ do not contribute to
the limit $W$.  The proof of this proposition can be found in
\cite{GMMP} and a formal derivation in \cite{RPK-WM}.

The initial conditions for the amplitudes $a_j$ are calculated using
the identity
\[
A_0^{-1}(\vx)=\dsum_{j\in J} \bb_j(\bx,\bk)\otimes\bb_j(\bx,\bk).
\]
Then (\ref{in-w}) implies that $a_{0}^{12}(0,\vx,\bk)=a_{0}^{21}(0,\vx,\bk)=0$
and
\begin{equation}
  \label{eq:in-a}
  a_{0}^{jj}(0,\vx,\bk)=a_\pm(0,\vx,\bk)=|\chi(\vx)|^2f(\bk),~~j=1,2.
\end{equation}

\subsection{Mode Decomposition and Refocusing}
\label{sec:modes}

We can use the above result to recast (\ref{eq:limitub}) as
\begin{equation}
  \label{eq:refocF}
  \bu^B(\Bxi;\bx_0)=(F(T,\cdot; \bx_0)*\bS)(\Bxi),
\end{equation}
where
\begin{eqnarray}
  &&F(T,\Bxi;\bx_0)=\dsum_{m,n=1}^2 \dint_{\Rm^\d}
   e^{i\bk\cdot\Bxi}
  a_0^{mn}(T,\bx_0;\bk) \Gamma \bb_m(\bx_0,\bk)\otimes\bb_n(\bx_0,\bk)
  A_0(\bx_0) \Gamma \frac{d\bk}{(2\pi)^\d}\nonumber\\
 \label{eq:F}
&&~~~~~~~~~~~~~~~
+\dint_{\Rm^\d}e^{i\bk\cdot\Bxi}
a_+(T,\bx_0;\bk) \Gamma \bb_+(\bx_0,\bk)\otimes\bb_+(\bx_0,\bk)
  A_0(\bx_0) \Gamma \frac{d\bk}{(2\pi)^\d}\\
&&~~~~~~~~~~~~~~~
+\dint_{\Rm^\d}e^{i\bk\cdot\Bxi}
a_-(T,\bx_0;\bk) \Gamma \bb_-(\bx_0,\bk)\otimes\bb_-(\bx_0,\bk)
  A_0(\bx_0) \Gamma \frac{d\bk}{(2\pi)^\d}.\nonumber
\end{eqnarray}
This expression can be used to assess the quality of the refocusing.
When $F(T,\Bxi;\bx_0)$ has a narrow support in $\Bxi$, refocusing is
good. When its support in $\Bxi$ grows larger, its quality degrades.
The spatial decay of the kernel $F(t,\vxi;\vx_0)$ in $\vxi$ is
directly related to the smoothness in $\bk$ of its Fourier transform
in $\vxi$:
\begin{eqnarray*}
&&\hat F(T,\bk;\vx_0)=\dsum_{m,n=1}^2
  a_0^{mn}(T,\bx_0;\bk) \Gamma \bb_m(\bx_0,\bk)\otimes\bb_n(\bx_0,\bk)
  A_0(\bx_0) \Gamma \frac{d\bk}{(2\pi)^\d}\\
&&
+\Gamma\left[
a_+(T,\bx_0;\bk)\bb_+(\bx_0,\bk)\otimes\bb_+(\bx_0,\bk)
\!+a_-(T,\bx_0;\bk) \bb_-(\bx_0,\bk)\otimes\bb_-(\bx_0,\bk)\right]
  A_0(\bx_0) \Gamma .
\end{eqnarray*}
Namely, for $F$ to decay in $\vxi$, one needs $\hat F(\bk)$ to be
smooth in $\bk$.  However, the eigenvectors $\vb_j$ are singular at
$\bk=0$ as can be seen from the explicit expressions (\ref{eq:bbpm}).
Therefore, $\hat F$ a priori is not smooth at $\bk=0$. This means that
in order to obtain good refocusing one needs the original signal to
have no low frequencies: $\hat S(\bk)=0$ near $\bk=0$. Low frequencies
in the initial data will not refocus well.

We can further simplify (\ref{eq:refocF})-(\ref{eq:F}) is we assume
that the initial source is irrotational. Taking Fourier transform of
both sides in (\ref{eq:refocF}), we obtain that
\begin{equation}\label{eq:ub-hat}
  \hat \bu^B(\bk;\bx_0)= \dsum_{j,n\in J}a_j(T,\bx_0,\bk)
\hat S_n(\bk)(A_0(\vx_0)\Gamma\vb_n(\vx_0,\bk)\cdot\vb_j(\vx_0,\bk))
\Gamma\vb_j(\vx_0,\bk)
\end{equation}
where we have defined
\begin{equation}
   \label{eq:decompS}
   \hat \bS(\bk)=\dsum_{n\in J} \hat S_n(\bk)\bb_n(\bx_0,\bk).
\end{equation}
Irrotationality of the initial source means that $\hat S_1$ and
$\hat S_2$ identically vanish, or equivalently that
\begin{equation}
    \label{eq:irrot-1}
  \bS(\vx)=\left(\begin{matrix}\nabla\phi(\vx)\cr p(\vx)\cr\end{matrix}\right)
\end{equation}
for some pressure $p(\bx)$ and potential $\phi(\bx)$.  Remarking that
$\Gamma\bb_\pm=-\bb_{\mp}$ and by irrotationality that
$(A_0(\bx_0)\hat\bS(\bk)\cdot\bb_{1,2}(\bk))=0$, we use
(\ref{eq:scalarA0}) to recast (\ref{eq:ub-hat}) as
\begin{equation}
   \label{eq:ub-hat1}
  \hat \bu^B(\bk;\bx_0)= a_-(T,\bx_0,\bk)
  \hat S_+(\bk)\bb_+(\bx_0,\bk) +a_+(T,\bx_0,\bk)
  \hat S_-(\bk)\bb_-(\bx_0,\bk).
\end{equation}
Decomposing the source $\bS(\vx)$ as
\begin{displaymath}
  \bS(\bx)=\bS_+(\bx)+\bS_-(\bx),\qquad\mbox{such that} \quad
  \hat \bS_\pm(\bk)=\hat S_\pm(\bk) \bb_\pm(\bx_0,\bk),
\end{displaymath}
the back-propagated signal takes the form
\begin{equation}
   \label{eq:ub-1}
   \bu^B(\vxi;\bx_0)= (\hat a_-(T,\vx_0,\cdot)*\bS_+(\cdot))(\vxi)+
   (\hat a_+(T,\vx_0,\cdot)*\bS_-(\cdot))(\vxi)
\end{equation}
where $\hat a_\pm$ is the Fourier of $a_\pm$ in $\bk$.  This form is
much more tractable than (\ref{eq:refocF})-(\ref{eq:F}). It is also
almost as general. Indeed, rotational modes do not propagate in the
high frequency regime. Therefore, they are exactly back-propagated
when $\chi(\bx_0)=1$ and $f(\bx)=\delta(\bx)$, and not back-propagated
at all when $\chi(\bx_0)=0$. All the refocusing properties are thus
captured by the amplitudes $a_\pm(T,\bx_0,\bk)$. Their evolution
equation characterizes how waves propagate in the medium and their
initial conditions characterize the recording array.

\subsection{Homogeneous Media}
\label{sec:homog}

In homogeneous media with $c(\bx)=c_0$ the amplitudes
$a_\pm(T,\vx,\bk)$ satisfy the free transport equation \cite{GMMP,RPK-WM}
\begin{equation}
  \label{eq:freetr}
  \pdr{a_\pm}{t}\pm c_0\vhatk\cdot\nabla_\vx a_\pm=0
\end{equation}
with initial data $a_\pm(0,\vx,\bk)=|\chi(\vx)|^2 f(\bk)$ as in
(\ref{eq:in-a}).  They are therefore given by
\begin{equation}
  \label{eq:a-uniform}
  a_\pm(t,\bx_0,\bk)=|\chi(\bx_0\mp c_0\hat \bk t)|^2 \hat f(\bk).
\end{equation}
These amplitudes become more and more singular in $\bk$ as time grows
since their gradient in $\bk$ grows linearly with time. The
corresponding kernel $F=F_H$ decays therefore more slowly in $\Bxi$ as
time grows. This implies that the quality of the refocusing degrades
with time. For sufficiently large times, all the energy has left the
domain $\Omega$ (assumed to be bounded), and the coefficients
$a_\pm(t,\bx_0,\bk)$ vanish.  Therefore the back-propagated signal
$\bu^B(\Bxi;\bx_0)$ also vanishes, which means that there is no
refocusing at all.  The same conclusions could also be drawn by
analyzing (\ref{eq:refoc}) directly in a homogeneous medium.  This is the
situation in the numerical experiment presented in Fig.
\ref{fig:num1}: in a homogeneous medium, the back-propagated signal would
vanish.

\subsection{Heterogeneous Media and Radiative Transport Regime}
\label{sec:rte}

The results of the preceding sections show how the back-propagated
signal $\bu^B(\Bxi;\bx_0)$ is related to the propagating modes
$a_\pm(T,\bx_0,\bk)$ of the Wigner matrix $W(T,\bx_0,\bk)$. The form
assumed by the modes $a_\pm(T,\bx_0,\bk)$, and in particular their
smoothness in $\bk$, will depend on the hypotheses we make on the
underlying medium; i.e., on the density $\rho(\vx)$ and
compressibility $\kappa(\vx)$ that appear in the matrix $A(\bx)$. We
have seen that partial measurements in homogeneous media yield poor
refocusing properties. We now show that refocusing is much better in
random media.

We consider here the radiative transport regime, also known as weak
coupling limit. There, the fluctuations in the physical parameters are
weak and vary on a scale comparable to the scale of the initial
source. Density and compressibility assume the form
\begin{equation}\label{eq:rte-scaling}
  \rho(\bx)=\rho_0 +\sqrt\eps\rho_1(\dfrac{\bx}{\eps})
  \qquad \mbox{ and } \qquad
  \kappa(\bx)=\kappa_0 + \sqrt\eps\kappa_1(\dfrac{\bx}{\eps}).
\end{equation}
The functions $\rho_1$ and $\kappa_1$ are assumed to be mean-zero
spatially homogeneous processes.  The average (with respect to
realizations of the medium) of the propagating amplitudes $a_\pm$,
denoted by $\bar a_\pm$, satisfy in the high frequency limit $\eps\to
0$ a radiative transfer equation (RTE), which is a linear Boltzmann
equation of the form
\begin{equation}
  \label{eq:rtea}
  \begin{array}{l}
   \pdr{\bar a_\pm}{t} \pm c_0 \hat \bk\cdot\nabla_\vx \bar a_\pm
= \dint_{\Rm^\d}
   \sigma(\bk,\bp) (\bar a_\pm(t,\bx,\bp)-\bar a_\pm(t,\bx,\bk))
     \delta(c_0(|\bk|-|\bp|)) d\bp \\
  \bar a_\pm(0,\bx,\bk)=|\chi(\bx)|^2\hat f(\bk).
  \end{array}\!\!
\end{equation}
The scattering coefficient $\sigma(\bk,\bp)$ depends on the power
spectra of $\rho_1$ and $\kappa_1$.  We refer to \cite{RPK-WM} for the
details of the derivation and explicit form of $\sigma(\bk,\bp)$. The
above result remains formal for the wave equation and requires to
average over the realizations of the random medium although this is
not necessary in the physical and numerical time reversal experiments.
A rigorous proof of the derivation of the linear Boltzmann equation
(which also requires to average over realizations) has only been
obtained for the Schr\"odinger equation; see \cite{Erdos-Yau2,Spohn}.
Nevertheless, the above result formally characterizes the filter
$F(T,\Bxi;\bx_0)$ introduced in (\ref{eq:F}) and
(\ref{eq:ub-1}).

The transport equation (\ref{eq:rtea}) has a smoothing effect best
seen in its integral formulation. Let us define the total scattering
coefficient
$\Sigma(\bk)=\int_{\Rm^\d}\sigma(\bk,\bp)\delta(c_0(|\bk|-|\bp|))d\bp$.
Then the transport equation (\ref{eq:rtea}) may be rewritten as
\begin{eqnarray}
\label{eq:integral-form}
&&\bar a_\pm(t,\vx,\bk)=\bar a_\pm(0,\vx\mp c_0\vhatk t,\bk)e^{-\Sigma(\bk)t}\\
&&+\frac{|\bk|^{2}}{c_0}\int_0^t ds \dint_{S^{2}}\sigma(\bk,|\bk|\vhatp)
\bar a_\pm(s,\bx\mp c_0(t-s)\vhatk,|\bk|\vhatp)
e^{-\Sigma(\bk)(t-s)}d\Omega(\vhatp).
\nonumber
\end{eqnarray}
Here $\vhatp=\bp/|\bp|$ is the unit vector in direction of $\bp$ and
$d\Omega(\vhatp)$ is the surface element on the sphere $S^{2}$. The
first term in (\ref{eq:integral-form}) is the ballistic part that
undergoes no scattering. It has no smoothing effect, and, moreover, if
$a(0,\vx,\bk)$ is not smooth in $\vx$, as may be the case for
(\ref{eq:in-a}), the discontinuities in $\vx$ translate into
discontinuities in $\bk$ at latter times as in (\ref{eq:a-uniform}) in
a homogeneous medium. However, in contrast to the homogeneous medium
case, the ballistic term decays exponentially in time, and does not
affect the refocused signal for sufficiently long times $t\gg
1/\Sigma$. The second term in (\ref{eq:integral-form}) exhibits a
smoothing effect. Namely the operator ${\cal L}g$ defined by
\[
{\cal L}g(t,\vx,\bk)=\frac{|\bk|^{2}}{c_0}
\int_0^t ds \dint_{S^{2}}\sigma(\bk,|\bk|\vhatp)
g(s,\bx\mp c_0(t-s)\vhatk,|\bk|\vhatp)
e^{-\Sigma(\bk)(t-s)}d\Omega(\vhatp)
\]
is regularizing, in the sense that the function $\tilde g={\cal L}g$
has at least 1/2-more derivatives than $g$ (in some Sobolev scale).
The precise formulation of this smoothing property is given by the
averaging lemmas \cite{glps,mokhtar} and will not be dwelt upon here.
Iterating (\ref{eq:integral-form}) $n$ times we obtain
\begin{equation}
\label{eq:iterated}
\bar a_\pm(t,\vx,\bk)=a_\pm^{0}(t,\vx,\bk)+a_\pm^1(t,\vx,\bk)+\dots+
a_\pm^n(t,\vx,\bk)+ {\cal L}^{n+1}\bar a_\pm(t,\vx,\bk).
\end{equation}
The terms $a_\pm^0, \dots,a_\pm^{n}$ are given by
\[
a_\pm^0(t,\vx,\bk)=\bar a_\pm(0,\vx\mp c_0\vhatk t,\bk)e^{-\Sigma(\bk)t},~~
a_\pm^j(t,\vx,\bk)={\cal L}a_\pm^{j-1}(t,\vx,\bk).
\]
They describe, respectively, the contributions from waves that do not
scatter, scatter once, twice, \ldots . It is straightforward to verify
that all these terms decay exponentially in time and are negligible
for times $t\gg 1/\Sigma$. The last term in (\ref{eq:iterated}) has at
least $n/2$ more derivatives than the initial data $a_0$, or the
solution (\ref{eq:a-uniform}) of the homogeneous transport equation.
This leads to a faster decay in $\vxi$ of the Fourier transforms $\hat
a_\pm(T,\vx_0,\vxi)$ of $a_\pm(T,\vx_0,\bk)$ in $\bk$.  This gives a
qualitative explanation as to why refocusing is better in
heterogeneous media than in homogeneous media. A more quantitative
answer requires to solve the transport equation (\ref{eq:rtea}).

\subsection{Diffusion Regime}
\label{sec:diffusion}

It is known for times $t$ much longer than the scattering mean free
time $\tau_{{sc}}=1/\Sigma$ and distances of propagation $L$ very
large compared to $l_{{sc}}=c_0\tau_{{sc}}$ that solutions to the
radiative transport equation (\ref{eq:rtea}) can be approximated by
solutions to a diffusion equation, provided that $c(\bx)=c_0$ is
independent of $\bx$ \cite{dlen6,LK}. More precisely, we let
$\delta=l_{sc}/L\ll 1$ be a small parameter and rescale time and space
variables as $t\to t/\delta^2$ and $\vx\to\vx/\delta$. In this limit,
wave direction is completely randomized so that
\[
\bar a_+(t,\vx,\bk)\approx \bar a_-(t,\vx,\bk)\approx a(t,\vx,|\bk|),
\]
where $a$ solves
\begin{equation}
  \label{eq:diff}
  \begin{array}{l}
  \pdr{a(t,\bx,|\bk|)}{t}-D(|\bk|)\Delta_\bx a(t,\bx,|\bk|)=0,\\
  a(0,\bx,|\bk|)=|\chi(\bx)|^2
   \dfrac{1}{4\pi|\bk|^{2}}\dint_{\Rm^\d}\hat f(\bq)\delta(|\bq|-|\bk|)
    d\bq.
  \end{array}
\end{equation}
The diffusion coefficient $D(|\bk|)$ may be expressed explicitly in
terms of the scattering coefficient $\sigma(\bk,\bp)$ and hence
related to the power spectra of $\rho_1$ and $\kappa_1$. We refer to
\cite{RPK-WM} for the details.  For instance, let us assume for
simplicity that the density is not fluctuating, $\rho_1\equiv0$, and
that the compressibility fluctuations are delta-correlated, so that
${\mathbb E}\{\hat\kappa_1(\bp)\hat\kappa_1(\bq)\}= \kappa_0^2 \hat
R_0\delta(\bp+\bq)$. Then we have
\begin{equation}\label{sigmas}
\sigma(\bk,\bp)=\dfrac{\pi c_0^2 |\bk|^2\hat R_0}{2},\qquad
\Sigma(|\bk|)={2\pi^2 c_0|\bk|^{4}\hat R_0}
\end{equation}
and
\begin{equation}\label{diff-coeff}
D(|\bk|)=\frac{c_0^2}{3\Sigma(|\bk|)}=\dfrac{c_0}{6\pi^2|\bk|^4\hat R_0}
\end{equation}

Let us assume that there are no initial rotational modes, so that the
source $\bS(\vx)$ is decomposed as in (\ref{eq:irrot-1}).
Using (\ref{eq:ub-hat1}), we
obtain that
\begin{equation}
  \label{eq:filtersimp}
  \hat \bu^B(\bk;\bx_0)= a(T,\bx_0,|\bk|) \hat \bS(\bk).
\end{equation}
When $f(\bx)$ is isotropic so that $\hat f(\bk)=\hat f(|\bk|)$, and
the diffusion coefficient is given by (\ref{diff-coeff}), the solution
of (\ref{eq:diff}) takes the form
\begin{equation}
  \label{eq:filterdiff}
  a(T,\bx_0,|\bk|)=\hat f(|\bk|)
\Big(\dfrac{3\pi|\bk|^4\hat R_0}{2c_0T}\Big)^{3/2}
\!\! \dint_{\Rm^\d}\!\!
  \exp\Big(-\dfrac{3\pi^2|\bk|^4\hat R_0
|\bx_0-\by|^2}{2c_0T}\Big)|\chi(\by)|^2 d\by.
\end{equation}
When $f(\vx)=\delta(\vx)$, and $\Omega=\Rm^\d$, so that
$\chi(\vx)\equiv1$, we retrieve $a(T,\bx_0,\bk)\equiv 1$, hence the
refocusing is perfect.  When only partial measurement is available,
the above formula indicates how the frequencies of the initial pulse
are filtered by the one-step time reversal process.  Notice that both
the low and high frequencies are damped. The reason is that low
frequencies scatter little with the underlying medium so that it takes
a long time for them to be randomized. High frequencies strongly
scatter with the underlying medium and consequently propagate little
so that the signal that reaches the recording array $\Omega$ is small
unless recorders are also located at the source point:
$\bx_0\in\Omega$. In the latter case they are very well measured and
back-propagated although this situation is not the most interesting
physically. Expression (\ref{eq:filterdiff}) may be generalized to
other power spectra of medium fluctuations in a straightforward manner
using the formula for the diffusion coefficient in \cite{RPK-WM}.

\subsection{Numerical Results}
\label{sec:num2}

The numerical results in Fig. \ref{fig:num1} show that some signal
refocuses at the location of the initial source after the time
reversal procedure. Based on the above theory however, we do not
expect the refocused signal to have exactly the same shape as the
original one. Since the location of the initial source belongs to the
recording array ($\chi(\bx_0)=1$) in our simulations, we expect from
our theory that high frequencies will refocus well but that low
frequencies will not.
\begin{figure}[htbp]
  \begin{center}
 $\mbox{}\!\!\!\!\!\!\!\!\!\!\!\!\!\!\!\!\!\!\!\!\!\!\!\!\!\!\!$
  \includegraphics[height=5.6cm]{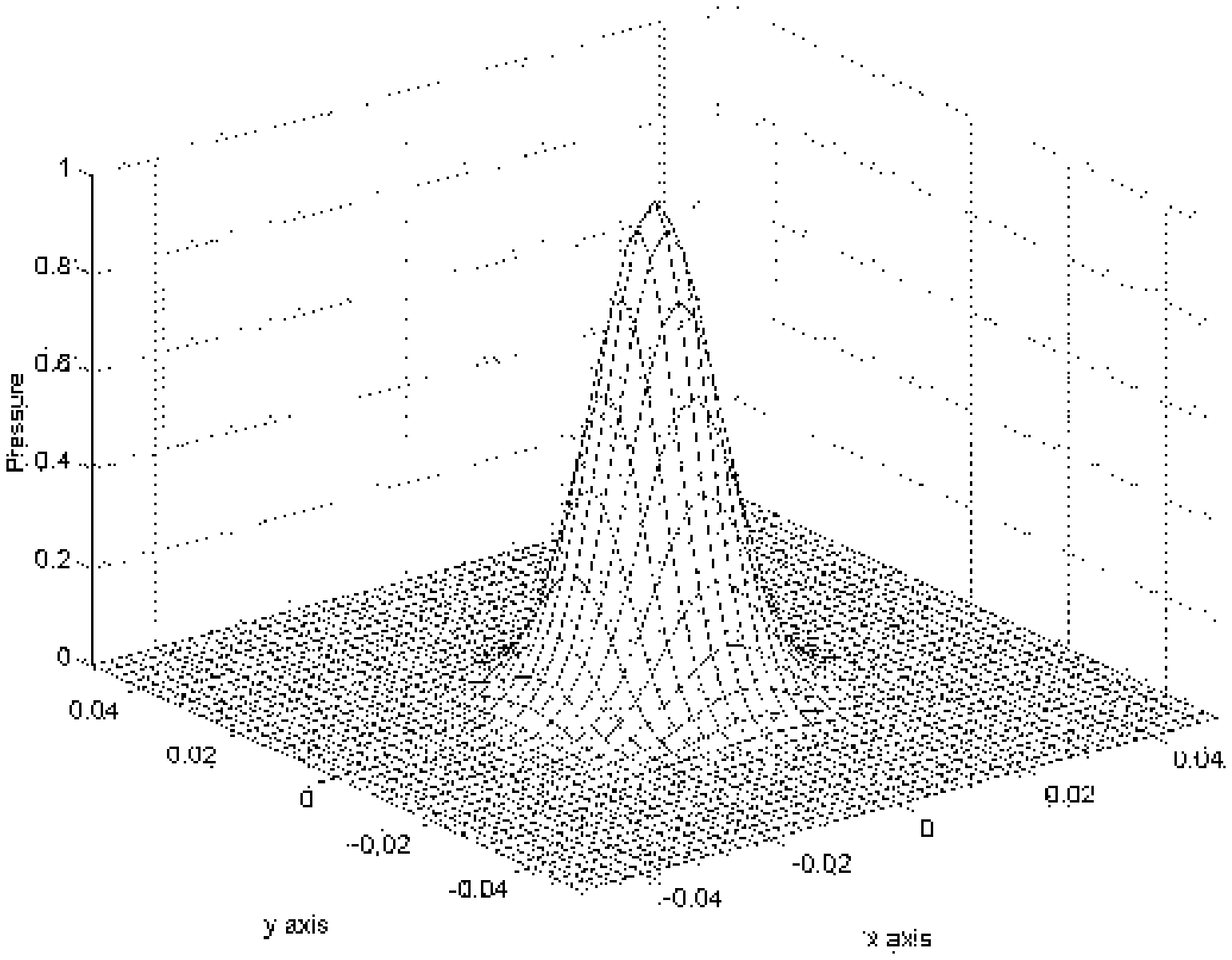}$\!\!\!\!\!\!\!\!\!\!\!\!$
  \includegraphics[height=5.6cm]{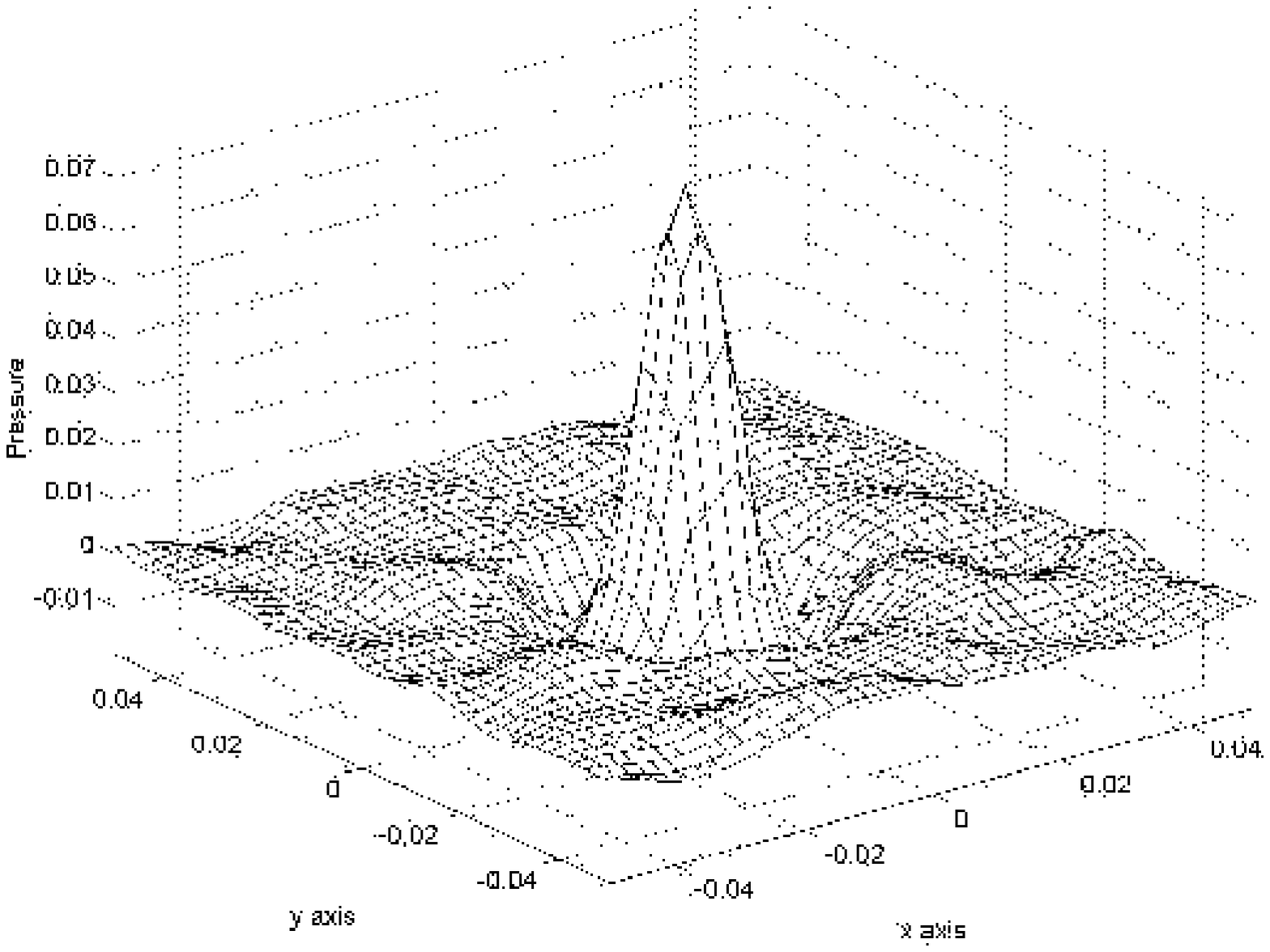}
    $\mbox{}\!\!\!\!\!\!\!\!\!\!\!\!\!\!\!\!\!\!\!\!\!\!\!\!\!\!\!$
    \caption{Zoom of the initial source and the refocused signal
    for the numerical experiment of Fig. \ref{fig:num1}.}
    \label{fig:num2}
  \end{center}
\end{figure}
This is confirmed by the numerical results in Fig. \ref{fig:num2},
where a zoom in the vicinity of $\bx_0={\bf 0}$ of the initial source
and refocused signal are represented. Notice that the numerical
simulations are presented here only to help in the understanding of the
refocusing theory and do not aim at reproducing the theory in a
quantitative manner. The random fluctuations are quite strong in our
numerical simulations and it is unlikely that the diffusive regime may
be valid. The refocused signal on the right figure looks however like
a high-pass filter of the signal on the left figure, as expected from
theory.

\section{Refocusing of Classical Waves}
\label{sec:ref}

The theory presented in section \ref{sec:TRRM} provides a quantitative
explanation for the results observed in time reversal physical and
numerical experiments.  However, the time reversal procedure is by no
means necessary to obtain refocusing. Time reversal is associated with
the specific choice (\ref{eq:gamma}) for the matrix $\Gamma$ in the
preceding section, which reverses the direction of the acoustic
velocity and keeps pressure unchanged. Other choices for $\Gamma$ are
however possible.  When nothing is done at time $T$, i.e., when we
choose $\Gamma=I$, no refocusing occurs as one might expect. It turns
out that $\Gamma=I$ is more or less the only choice of a matrix that
prevents some sort of refocusing. Section \ref{sec:refth} presents the
theory of refocusing for acoustic waves, which is corroborated by
numerical results presented in section \ref{sec:num3}.  Sections
\ref{sec:galwaves} and \ref{sec:diff-2} generalize the theory to other
linear hyperbolic systems.

\subsection{General Refocusing of Acoustic Waves}
\label{sec:refth}

In one-step time reversal, the action of the ``intelligent'' array is
captured by the choice of the signal processing matrix $\Gamma$ in
(\ref{eq:prop}).  Time reversal is characterized by $\Gamma$ given in
(\ref{eq:gamma}).  A passive array is characterized by
$\Gamma=I$. This section analyzes the role of other choices for
$\Gamma$, which we let depend on the receiver location so that each receiver
may perform its own kind of signal processing.

The signal after time reversal is still given by (\ref{eq:prop}),
where $\Gamma(\by')$ is now arbitrary. At time $2T$, after
back-propagation, we are free to multiply the signal by an arbitrary
invertible matrix to analyze the signal. It is convenient to multiply
the back-propagated signal by the matrix $\Gamma_0={\rm
Diag}(-1,-1,-1,1)$ as in classical time reversal.  The reconstruction
formula (\ref{eq:refoceps}) in the localized source limit is then
replaced by
\begin{equation}
  \label{eq:refocepsbis}
   \bu^B(\Bxi;\bx_0)=\dint_{\Rm^{9}}
    \Gamma_0 G(T,\bx_0+\eps\Bxi;\by)
       \Gamma(\by') G(T,\by';\bx_0+\eps \bz)
    \chi(\by,\by')  \bS(\bz)
    d\by d\by' d\bz
\end{equation}
with $\chi(\vy,\vy')$ defined by (\ref{chieps}).  To generalize the
results of section \ref{sec:TRRM}, we need to define an appropriate
adjoint Green's matrix $G_*$. As before, this will allow us to remove
the matrix $\Gamma$ between the two Green's matrices in
(\ref{eq:refocepsbis}) and to interchange the order of points in the
second Green's matrix.  We define the new adjoint Green's function
$G_*(t,\bx;\by)$ as the solution to
\begin{equation}
  \label{eq:adjointGreenbis}
  \begin{array}{l}
  \pdr{G_*(t,\bx;\by)}{t} A(\bx) + \pdr{G_*(t,\bx;\by)}{x^j}D^j =0 \\
\\
  G_*(0,\bx;\by)=\Gamma(\bx)\Gamma_0 A^{-1}(\bx)\delta(\bx-\by).
  \end{array}
\end{equation}
Following the steps of section \ref{sec:adjoint}, we show that
\begin{equation}
  \label{eq:relgreenbis}
  G_*(t,\bx,\by)=\Gamma(\by) G(t,\by;\bx)A^{-1}(\bx) \Gamma_0.
\end{equation}
The only modification compared to the corresponding derivation of
(\ref{eq:relgreen}) is to multiply (\ref{eq:Giny}) on the left by
$\Gamma(\bx)$ and on the right by $\Gamma_0$ so that $\Gamma(\by)$
appears on the left in (\ref{eq:GtoG*}). The re-transmitted signal may
now be recast as
\begin{eqnarray}
  \label{eq:refocepsbis3} 
\bu^B(\Bxi;\bx_0)&=&\dint_{\Rm^{9}}d\by d\by' d\bz
\Gamma_0 G(T,\bx_0+\eps\Bxi;\by)G_*(T,\bx_0+\eps \bz;\by')\Gamma_0^{-1}\\
&\times&  A(\bx_0+\eps\bz) \chi(\by,\by') \bS(\bz).\nonumber
\end{eqnarray}
Therefore the only modification in the expression for the
re-transmitted signal compared to the time reversed signal
(\ref{eq:refoceps2}) is in the initial data for
(\ref{eq:adjointGreenbis}), which is the only place where the matrix
$\Gamma(\vx)$ appears.

The analysis in sections \ref{sec:adjoint}-\ref{sec:rte} requires only
minor changes, which we now outline.  The back-propagated signal may
still be expressed in term of the Wigner distribution (compare to
(\ref{eq:refoceps4}))
\begin{equation}
  \label{eq:refoceps5}
   \bu^B(\Bxi;\bx_0)=\dint_{\Rm^{6}}e^{i\bk\cdot(\Bxi-\bz)}
\Gamma_0 W_\eps(T,\bx_0+\eps\dfrac{\bz+\Bxi}{2},\bk) \Gamma_0 A(\bx_0+\eps\bz)
      \bS(\bz) \frac{d\bz d\bk}{(2\pi)^\d}.
\end{equation}
The Wigner distribution is defined as before by (\ref{eq:Wigner}) and
(\ref{eq:WignerU}). The function $Q$ is defined as before as the
solution of (\ref{eq:direct}) with initial data
$Q(0,x;\bq)=\chi(\vx)e^{i\bq\cdot\vx/\eps}I$, while $Q_*$ solves
(\ref{eq:adjointGreen}) with the initial data
$Q_*(0,\vx;\bq)=\Gamma(\vx)\Gamma_0A^{-1}(\vx)\chi(\vx)e^{-i\bq\cdot\vx/\eps}$.
The initial Wigner distribution is now given by
\begin{equation}
  \label{eq:initWbis}
  W(0,\bx,\bk)= |\chi(\bx)|^2
   \Gamma(\bx) \Gamma_0 A^{-1}(\bx) \hat f(\bk).
\end{equation}
Lemma \ref{lem:bound} and Proposition \ref{ref:prop} also hold, and we
obtain the analog of (\ref{eq:limitub})
\begin{equation}
    \label{eq:limitubbis}
    \bu(\Bxi;\bx_0)=\dint_{\Rm^{6}} e^{i\bk\cdot(\Bxi-\bz)}
     \Gamma_0 W(T,\bx_0,\bk) \Gamma_0 A_0(\bx_0) \bS(\bz) d\bz d\bk.
\end{equation}
The limit Wigner distribution $W(T,\bx_0,\bk)$ admits the mode
decomposition (\ref{eq:decomp}) as before. If we assume that the
source $\bS(\vx)$ has the form (\ref{eq:irrot-1}) so that no
rotational modes are present initially, we recover the refocalization
formula (\ref{eq:ub-hat1}):
\begin{equation}\label{eq:ub-hat2}
  \hat \bu^B(\bk;\bx_0)= a_-(T,\bx_0,\bk)
\hat S_+(\bk)\vb_+(\vx_0,\bk) +a_+(T,\bx_0,\bk)
\hat S_-(\bk)\vb_-(\vx_0,\bk).
\end{equation}
The initial conditions for the amplitudes $a_\pm$ are
replaced by
\begin{eqnarray}
  \label{eq:initabis}
&&  a_\pm(0,\bx,\bk)=\hbox{Tr}\left[A_0(\vx)W(0,\vx,\bk)A_0(\vx)
\vb_\pm(\vx_0,\bk)\vb_\pm^*(\vx_0,\bk)\right]\\
&&=|\chi(\bx)|^2\hat f(\bk)
  (A_0(\vx)\Gamma(\vx)\vb_{\mp}(\bx,\bk)\cdot\vb_\pm(\bx,\bk)).\nonumber
\end{eqnarray}
Observe that when $\Gamma(\vx)=\Gamma_0$, we get back the results of
section \ref{sec:rte}.  When the signal is not changed at the array,
so that $\Gamma=I$, the coefficients $a_\pm(0,\vx,\bk)\equiv0$ by
orthogonality (\ref{eq:scalarA0}) of the eigenvectors $\bb_j$. We thus
obtain that no refocusing occurs when the ``intelligent'' array is
replaced by a passive array, as expected physically.

Another interesting example is when only pressure $p$ is measured, so
that the matrix $\Gamma=\hbox{Diag}(0,0,0,1)$. Then the initial data
is
\[
a_\pm(0,\vx,\bk)=\frac 12|\chi(\bx)|^2\hat f(\bk),
\]
which differs by a factor $1/2$ from the full time reversal case
(\ref{eq:in-a}). Therefore the re-transmitted signal $\bu^B$ also
differs only by a factor $1/2$ from the latter case, and the quality
of refocusing as well as the shape of the re-propagated signal are
exactly the same. The same observation applies to the measurement and
reversal of the acoustic velocity only, which corresponds to the
matrix $\Gamma=\hbox{Diag}(-1,-1,-1,0)$. The factor $1/2$ comes from
the fact that only the potential energy or the kinetic energy is
measured in the first and second cases, respectively.  For high
frequency acoustic waves, the potential and kinetic energies are
equal, hence the factor $1/2$. We can also verify that when only the
first component of the velocity field is measured so that
$\Gamma=\rm{Diag}(-1,0,0,0)$, the initial data is
\begin{equation}\label{in-first-comp}
a_\pm(0,\bx,\bk)=|\chi(\bx)|^2\hat f(\bk)\frac{k_1^2}{2|\bk|^2}.
\end{equation}

As in the time reversal setting of section \ref{sec:TRRM}, the quality
of the refocusing is related to the smoothness of the amplitudes
$a_\pm$ in $\bk$. In a homogeneous medium they satisfy the free
transport equation (\ref{eq:freetr}), and are given by
\begin{eqnarray*}
a_\pm(t,\bx,\bk)\!&=&\!|\chi(\bx-c_0\vhatk t)|^2\hat f(\bk)\\
&\times &(A_0(\vx-c_0\vhatk t)\Gamma(\vx-c_0\vhatk t)
\vb_{\mp}(\bx-c_0\vhatk t,\bk)\cdot\vb_\pm(\bx-c_0\vhatk t,\bk)).
\end{eqnarray*}
Once again, we observe that in a uniform medium $a_\pm$ become less
regular in $\bk$ as time grows, thus refocusing is poor.

The considerations of section \ref{sec:rte} show that in the radiative
transport regime the amplitudes $a_\pm$ become smoother in $\bk$ also
with initial data given by (\ref{eq:initabis}). This leads to a better
refocusing as explained in section \ref{sec:modes}.  Let us assume
that the diffusion regime of section \ref{sec:diffusion} is valid and
that the kernel $f$ is isotropic $\hat f(\bk)= \hat f(|\bk|)$. This
requires in particular that $A_0(\bx)$ be independent of $\bx$. We
obtain that $a_\pm(T,\bx_0,\bk)=\tilde a(T,\bx_0,|\bk|)$, thus the
refocusing formula (\ref{eq:ub-hat2}) reduces to
\begin{equation}
  \label{eq:refocbis}
  \hat \bu^B(\bk;\bx_0)= \tilde a(T,\bx_0,|\bk|) \hat \bS(\bk).
\end{equation}
The difference with the case treated in section \ref{sec:diffusion} is
that $\tilde a(T,\bx,|\bk|)$ solves the diffusion equation
(\ref{eq:diff}) with new initial conditions given by
\begin{eqnarray}
  \label{eq:icdiffbis}
  \tilde a(0,\bx,|\bk|)&=& \dfrac{|\chi(\bx)|^2}{4\pi|\bk|^{2}}
\dint_{\Rm^\d}\hat f(|\bq|)
(A_0\Gamma(\bx)\bb_-(\bq)\cdot\bb_+(\bq))
    \delta(|\bq|-|\bk|)d\bq\\
&=&
\dfrac{|\chi(\bx)|^2}{4\pi|\bk|^{2}}
\dint_{\Rm^\d}\hat f(|\bq|)
(A_0\Gamma(\bx)\bb_+(\bq)\cdot\bb_-(\bq))
    \delta(|\bq|-|\bk|)d\bq.\nonumber
\end{eqnarray}

When only the first component of the velocity field is measured, as in
(\ref{in-first-comp}), the initial data for $\tilde a$ is
\[
\tilde a(0,\vx,|\bk|)=\frac{1}{6}|\chi(\vx)|^2\hat f(|\bk|).
\]
Therefore even time reversing only one component of the acoustic velocity
field produces a re-propagated signal that is equal to the full
re-propagated field up to a constant factor.

More generally, we deduce from (\ref{eq:icdiffbis}) that a detector at
$\bx$ will contribute some refocusing for waves with wavenumber
$|\bk|$ provided that
\begin{displaymath}
  \dint_{S^{2}}
\hat f(|\bk|\hat\bq)
       (A_0\Gamma(\bx)\bb_\mp(\hat \bq)\cdot \bb_\pm(\hat\bq))
     d\Omega(\hat\bq) \not=0.
\end{displaymath}
When $f(\bx)=f(|\bx|)$ is radial, this property becomes independent of
the wavenumber $|\bk|$ and reduces to
\begin{math}
  \int_{S^{2}}
  (A_0\Gamma(\bx)\bb_\mp(\hat \bq)\cdot
\bb_\pm(\hat\bq)) d\Omega(\hat\bq) \not=0.
\end{math}

\subsection{Numerical Results}
\label{sec:num3}

\begin{figure}[htbp]
  \begin{center}
 $\mbox{}\!\!\!\!\!\!\!\!\!\!\!\!\!\!\!\!\!\!\!\!\!\!\!\!\!\!\!$
  \includegraphics[height=5.6cm]{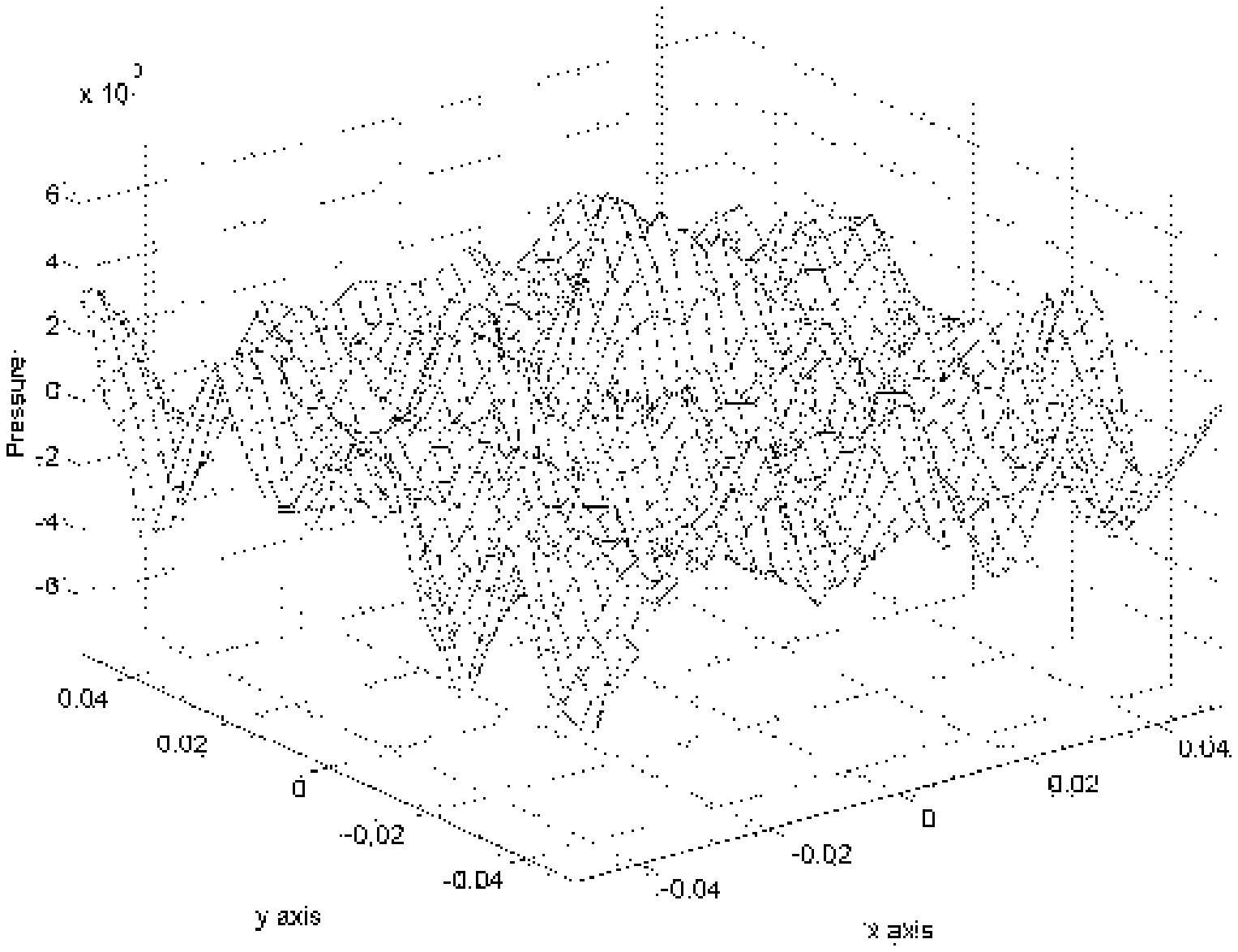}
    $\!\!\!\!\!\!\!\!\!\!\!\!$
  \includegraphics[height=5.6cm]{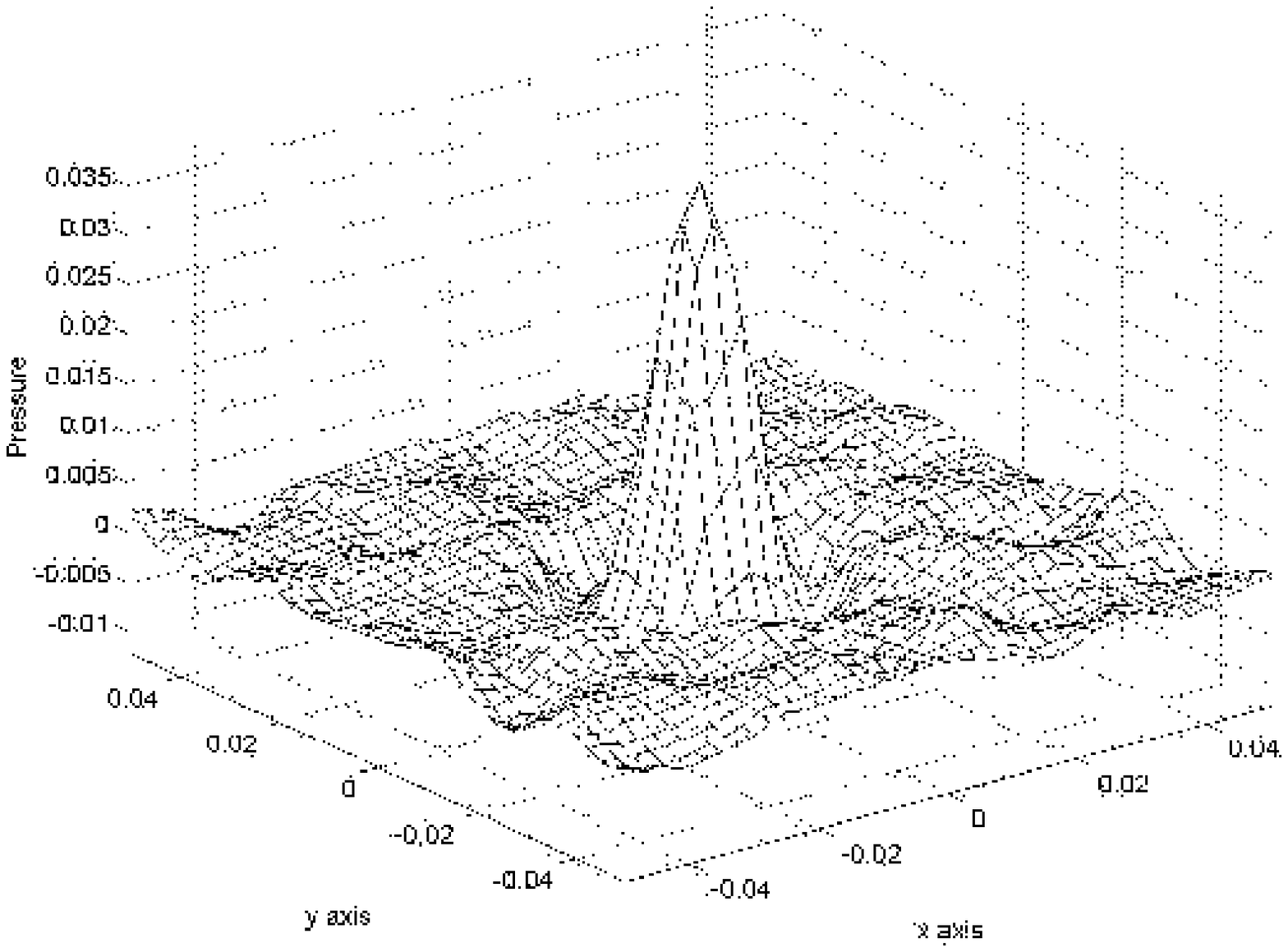}
    $\mbox{}\!\!\!\!\!\!\!\!\!\!\!\!\!\!\!\!\!\!\!\!\!\!\!\!\!\!\!$
    \caption{Zoom of the refocused signals
      for the numerical experiment of Fig. \ref{fig:num1} with
      processing $\Gamma=I$ (left), with a maximal amplitude of
      roughly $4\,10^{-3}$ and $\Gamma=\hbox{\rm Diag}(0,0,0,1)$ (right),
      with a maximal amplitude of roughly $0.035$.}
    \label{fig:num3}
  \end{center}
\end{figure}

Let us come back to the numerical results presented in Fig.
\ref{fig:num1} and \ref{fig:num2}. We now consider two different
processings at the recording array. The first array is passive,
corresponding to $\Gamma=I$, and the second array only measures
pressure so that $\Gamma=\hbox{Diag}(0,0,0,1)$. The zoom in the
vicinity of $\bx_0={\bf 0}$ of the ``refocused'' signals is given in
Fig. \ref{fig:num3}.
The left figure shows no refocusing, in accordance with physical
intuition and theory. The right figure shows that refocusing indeed
occurs when only pressure in recorded (and its time derivative is set
to $0$ in the solution of the wave equation presented in the
appendix). Notice also that the refocused signal is roughly one half
the one obtained in Fig. \ref{fig:num2} as predicted by theory.

\subsection{Refocusing of Other Classical Waves}
\label{sec:galwaves}

The preceding sections deal with the refocusing of acoustic waves.
The theory can however be extended to more complicated linear
hyperbolic systems of the form (\ref{eq:direct}) with $A(\vx)$ a
positive definite matrix, $D^j$ symmetric matrices, and
$\vu\in{\mathbb C}^m$.  These include electromagnetic and elastic
waves. Their explicit representation in the form (\ref{eq:direct}) and
expressions for the matrices $A(\bx)$ and $D^j$ in these cases may be
found in \cite{RPK-WM}.  For instance, the Maxwell equations
\begin{eqnarray*}
\pdr {\vE}t&=&\frac 1{\epsilon(\bx) }\hbox{  curl }\vH  \\
\pdr {\vH}t&=&-\frac 1{\mu(\bx) }\hbox{curl }\vE \nonumber
\end{eqnarray*}
may be written in the form (\ref{eq:direct}) with
$\vu=(\vE,\vH)\in{\mathbb C}^6$ and the matrix
$A(\bx)=\hbox{Diag}(\epsilon(\bx),\epsilon(\bx),\epsilon(\bx),\mu(\bx),\mu(\bx),\mu(\bx))$.
Here $\epsilon(\bx)$ is the dielectric constant (not to be confused
with the small parameter $\eps$), and $\mu(\bx)$ is the magnetic
permeability. The $6\times 6$ dispersion matrix $L(\bx,\bk)$ for the
Maxwell equations is given by
\begin{eqnarray*}
L(\bx,\bk)=-\left( \begin{matrix}
 0 & 0 & 0 & 0 & -k_3/\epsilon(\bx) & k_2/\epsilon(\bx)\cr
0 & 0 & 0 & k_3/\epsilon(\bx) & 0 & -k_1/\epsilon(\bx)\cr
0 & 0 & 0 & -k_2/\epsilon(\bx) & k_1/\epsilon(\bx) & 0\cr
0 & k_3/\mu(\bx) & -k_2/\mu(\bx) & 0 & 0 & 0 \cr
-k_3/\mu(\bx) & 0 & k_1/\mu(\bx) & 0 & 0 & 0 \cr
k_2/\mu(\bx) & -k_1/\mu(\bx) & 0 & 0 & 0 & 0 \cr
\end{matrix} \right).
\end{eqnarray*}

Generalization of our results for acoustic waves to such general
systems is quite straightforward so we concentrate only on the
modifications that need be made. The time reversal procedure is
exactly the same as before: a signal propagates from a localized
source, is recorded, processed as in (\ref{eq:prop}) with a general
matrix $\Gamma(\vy')$, and re-emitted into the medium.  The
re-transmitted signal is given by (\ref{eq:refocepsbis}).
Furthermore, the equation for the adjoint Green's matrix
(\ref{eq:adjointGreenbis}), the definition of the Wigner transform in
section \ref{sec:wigner}, and the expression (\ref{eq:limitubbis}) for
the re-propagated signal still hold.

The analysis of the re-propagated signal is reduced to the study of
the Wigner distribution, which is now modified.  The mode decomposition
need be generalized. We recall that
\begin{displaymath}
  L(\bx,\bk)=A_0^{-1}(\bx)k_j D^j
\end{displaymath}
is the $m\times m$ dispersion matrix associated with the hyperbolic
system (\ref{eq:direct}). Since $L(\bx,\bk)$ is symmetric with respect
to the inner product $\langle\bu,\bv\rangle_{A_0}= (A_0\bu\cdot\bv)$,
its eigenvalues are real and its eigenvectors form a basis.  We assume
the existence of a time reversal matrix $\Gamma_0$ such that
(\ref{eq:commutgamma}) holds with $\Gamma=\Gamma_0$ and such that
$\Gamma_0^2=I$. For example, for electromagnetic waves
$\Gamma_0=\hbox{Diag}(1,1,1,-1,-1,-1)$.  Then the spectrum of $L$ is
symmetric about zero and the eigenvalues $\pm\omega^\alpha$ have the
same multiplicity.  We assume in addition that $L$ is isotropic so
that its eigenvalues have the form $\omega_\pm^\alpha(\vx,\bk)=\pm
c^\alpha(\vx)|\bk|$, where $c_\alpha(\bx)$ is the speed of mode
$\alpha$. We denote by $r_\alpha$ their respective multiplicities,
assumed to be independent of $\bx$ and $\bk$ for $\bk\not=0$. The
matrix $L$ has a basis of eigenvectors $\bb_\pm^{\alpha,j}(\bx,\bk)$
such that
\begin{displaymath}
    L(\bx,\bk)\bb_\pm^{\alpha,j}(\bx,\bk)=
\pm \omega^\alpha(\vx,\bk)\bb_\pm^{\alpha,j}(\bx,\bk),
           \quad j=1,\ldots,r_\alpha, 
\end{displaymath}
and $\vb_\pm^{\alpha,j}$ form an orthonormal set with respect to the
inner product $\langle,\rangle_{A_0}$.  The different $\omega_\alpha$
correspond to different types of waves (modes).  Various indices
$1\leq j \leq r_\alpha$ refer to different polarizations of a given
mode.  The eigenvectors $\vb_+^{\alpha,j}$ and $\vb_-^{\alpha,j}$ are
related by
\begin{equation}
  \label{eq:relspectgamma0}
  \Gamma_0\bb_+^{\alpha,j}(\bx,\bk)=\bb_{-}^{\alpha,j}(\bx,\bk),~~
\Gamma_0\bb_-^{\alpha,j}(\bx,\bk)=\bb_{+}^{\alpha,j}(\bx,\bk).
\end{equation}
Proposition \ref{prop:decomp}
is then generalized as follows \cite{GMMP,RPK-WM}:
\begin{proposition}
  \label{prop:decompgal} There exist scalar functions
  $a_\pm^{\alpha,jm}(t,\bx,\bk)$ such that
\begin{equation}\label{eq:decompbis}
W(t,\bx,\bk)=\dsum_{\pm,\alpha,j,m}
  a_\pm^{\alpha,jm}(t,\bx,\bk)\bb_\pm^{\alpha,j}(\bx,\bk)
  \otimes\bb_\pm^{\alpha,m}(\bx,\bk).
\end{equation}
Here the sum runs over all possible values of $\pm$, $\alpha$, and
$1\leq j,m\leq r_\alpha$.
\end{proposition}

The main content of this proposition is again that the cross terms
$\bb_\pm^{\alpha,j}(\bx,\bk)\otimes\bb_\mp^{\beta,m}(\bx,\bk)$ do not
contribute, as well as the terms
$\bb_\pm^{\alpha,j}(\bx,\bk)\otimes\bb_\pm^{\alpha',m}(\bx,\bk)$ when
$\alpha\not=\alpha'$. This is because modes propagating with different
speeds do not interfere constructively in the high frequency limit.

We may now insert expression (\ref{eq:decompbis}) into
(\ref{eq:limitubbis}) and obtain the following generalization
of (\ref{eq:ub-hat2})
\begin{eqnarray}\label{eq:refoclimgal}
&&  \hat \bu^B(\bk;\bx_0)= \dsum_{\alpha,j,m}\left[
a_{-}^{\alpha,mj}(T,\bx_0,\bk)
\hat S_+^{\alpha,j}(\vx_0,\bk)\bb_+^{\alpha,m}(\bx_0,\bk)\right.\\
&&~~~~~~~~~~~~~~~+\left.
a_{+}^{\alpha,mj}(T,\bx_0,\bk)
\hat S_-^{\alpha,j}(\vx_0,\bk)\bb_-^{\alpha,m}(\bx_0,\bk)\right],\nonumber
\end{eqnarray}
where
\begin{math}
  \hat S_\pm^{\alpha,j}(\bk)=
(A(\vx_0)\hat \bS(\bk)\cdot\bb_\pm^{\alpha,j}(\bx_0,\bk)).
\end{math}
This formula tells us that only the modes that are present in the
initial source ($\hat S_\pm^{\alpha,j}(\bk)\not=0$) will be present in the
back-propagated signal but possibly with a different polarization, that is,
$j\ne m$.

The initial conditions for the modes
$a_{\pm}^{\alpha,jm}$ are given by
\begin{equation}
  \label{eq:initagal}
  a_\pm^{\alpha,jm}(0,\bx,\bk)=|\chi(\bx)|^2 \hat f(\bk)
     (A(\vx)\Gamma(\bx)\bb_{\mp}^{\alpha,m}(\bx,\bk)\cdot
     \bb_\pm^{\alpha,j}(\bx,\bk)),
\end{equation}
which generalizes (\ref{eq:initabis}). When $\Gamma(\bx)\equiv I$, we
again obtain that $a_\pm^{\alpha,jm}(0,\bx,\bk)\equiv0$, i.e., there is no
refocusing as physically expected. When $\Gamma(\bx)\equiv\Gamma_0$,
we have for all $\alpha$ that
\begin{displaymath}
  a_\pm^{\alpha,jm}(0,\bx,\bk)=|\chi(\bx)|^2 \hat f(\bk)\delta_{jm}.
\end{displaymath}
In a uniform medium the amplitudes $a_\pm^{\alpha,jm}$ satisfy an uncoupled
system of free transport equations (\ref{eq:freetr}):
\begin{equation}\label{eq:freetr-2}
\pdr{a_\pm^{\alpha,jm}}{t}\pm c_\alpha\vhatk\cdot\nabla_\vx a_\pm^{\alpha,jm}=0,
\end{equation}
which have no smoothing effect, and hence refocusing in a homogeneous
medium is still poor. When $f(\bx)=\delta(\bx)$ and $\Omega=\Rm^\d$, so
that $\chi(\vx)\equiv 1$, we still have that
$a_\pm^{\alpha,jm}(T,\bx_0,\bk)=\delta_{jm}$ and refocusing is again
perfect, that is, $\vu^B(\vxi;\vx_0)=\bS(\vxi)$, as may be seen from
(\ref{eq:refoclimgal}).

\subsection{The diffusive regime}\label{sec:diff-2}
The radiative transport regime holds when the matrices $A(\vx)$ have
the form
\[
A(\vx)=A_0(\vx)+\sqrt{\eps}A_1\Big(\dfrac{\bx}{\eps}\Big),
\]
as in (\ref{eq:rte-scaling}). Then the $r_\alpha\times r_\alpha$
coherence matrices $w_\pm^\alpha$ with entries $w_{\pm,jm}^{\alpha}=
a_\pm^{\alpha,jm}$ satisfy a system of matrix-valued radiative
transport equations (see \cite{RPK-WM} for the details) similar to
(\ref{eq:rtea}).
The matrix transport equations simplify considerably in the diffusive
regime, such as the one considered in section \ref{sec:diffusion} when
waves propagate over large distances and long times. We assume for
simplicity that $A_0=A_0(\bx)$ and $\Gamma=\Gamma(\bx)$ are
independent of $\bx$.  Polarization is lost in this regime, that is,
$a^{\alpha,jm}(t,\bx,\bk)=0$ for $j\not=m$ and wave energy is
equidistributed over all directions. This implies that
\[
a_+^{\alpha,jj}(t,\vx,\bk)=a_-^{\alpha,jj}(t,\vx,\bk)
=a_\alpha(t,\bx,|\bk|)
\]
so that $a^{\alpha,jj}$ is independent of $j=1,\dots,r_\alpha$ and of
the direction $\vhatk=\bk/|\bk|$.  Furthermore, because of multiple
scattering, a universal equipartition regime takes place so that
\begin{equation}
  \label{eq:equip}
  a_\alpha(t,\bx_0,|\bk|)=\phi(t,\bx_0,c_\alpha|\bk|),
\end{equation}
where $\phi(t,\bx,\omega)$ solves a diffusion equation in $\bx$ like
(\ref{eq:diff}) (see \cite{RPK-WM}). The diffusion coefficient
$D(\omega)$ may be expressed explicitly in terms of the power spectra
of the medium fluctuations \cite{RPK-WM}.  Using (\ref{eq:initagal})
and (\ref{eq:equip}), we obtain when $f$ is isotropic the following
initial data for the function $\phi$
\begin{equation}
  \label{eq:initphi}
  \phi(0,\bx,\omega)=\dfrac1{4\pi} |\chi(\bx)|^2
   \dint_{S^2}
   \dfrac{2}{|\alpha|}\dsum_{j,\omega_\alpha>0}
   \hat f\Big(\dfrac{\omega}{c_\alpha}\Big)
(A_0\Gamma \bb_-^{\alpha,j}(\hat\bk),\bb_+^{\alpha,j}(\hat\bk))
d\Omega(\hat\bk),
\end{equation}
where $|\alpha|$ is the number of non-vanishing eigenvalues of $L(\bx,\bk)$,
and $d\Omega(\hat\bk)$ is the Lebesgue measure on the unit sphere $S^2$.

Let us assume that non-propagating modes are absent in the initial
source $\bS(\vx)$, that is, $\hat S_0^j(\bk)=0$ with the subscript
zero referring to modes corresponding to $\omega_0=0$.
Then (\ref{eq:refoclimgal}) becomes
\begin{equation}
\label{eq:refoclimgal2}
\hat \bu(\bk;\bx_0)= \dsum_{\alpha,j}\phi(T,\bx_0,c_\alpha|\bk|)
\left[\hat S_+^{\alpha,j}(\bk)\bb_+^{\alpha,j}(\bx_0,\bk)+
\hat S_-^{\alpha,j}(\bk)\bb_-^{\alpha,j}(\bx_0,\bk)\right].
\end{equation}
This is an explicit expression for the re-propagated signal in the
diffusive regime, where $\phi$ solves the diffusion equation
(\ref{eq:diff}) with initial conditions (\ref{eq:initphi}).

\section{Conclusions}

This paper presents a theory that quantitatively describes the
refocusing phenomena in time reversal acoustics as well as for more
general processings of other classical waves. We show that the
back-propagated signal may be expressed as the convolution
(\ref{eq:refocintro}) of the original source $\bS$ with a filter $F$.
The quality of the refocusing is therefore determined by the spatial
decay of the kernel $F$. For acoustic waves, the explicit expression
(\ref{eq:F}) relates $F$ to the Wigner distribution of certain
solutions of the wave equation. The decay of $F$ is related to the
smoothness in the phase space of the amplitudes $a_j(t,\vx,\bk)$
defined in Proposition \ref{prop:decomp}. The latter satisfy a free
transport equation in homogeneous media, which sharpens the gradients
of $a_j$ and leads to poor refocusing. In contrast, the amplitudes
$a_j$ satisfy the radiative transport equation (\ref{eq:rtea}) in
heterogeneous media, which has a smoothing effect. This leads to a
rapid spatial decay of the filter $F$ and a better refocusing.  For
longer times, $a_j$ satisfies a diffusion equation. This allows for an
explicit expression (\ref{eq:filtersimp})-(\ref{eq:filterdiff}) of the
time reversed signal. The same theory holds for more general waves
and more general processing procedures at the recording array, which
allows us to describe the refocusing of electromagnetic waves when
only one component of the electric field is measured, for instance.

\section*{Appendix}
This appendix presents the details of the numerical simulation of
(\ref{eq:wave}). We assume that $\rho$ is constant and that only
$\kappa(\bx)$ fluctuates. We can therefore recast (\ref{eq:wave}) as
\begin{displaymath}
  \pdrt{p}{t} - c^2(\bx) \Delta p=0.
\end{displaymath}
The above wave equation is discretized using a second-order scheme
(three point stencil in every variable) both in time and space. The
resolution in time is explicit and time reversible, i.e., the equation
that yields $p(t_{n+1})$ from $p(t_{n-1})$ and $p(t_{n})$ can be used
to retrieve $p(t_{n-1})$ exactly from $p(t_{n})$ and $p(t_{n+1})$. We
write $c^2(\bx)=c_0^2+c_1^2(\bx)$.  The average velocity is $c_0^2=1$.
The random part $c_1^2$ has been constructed as follows. Let $2N\times
2N$ be the number of spatial grid points and $c_{1;n,m}^2$ be the
value of $c_1^2$ at the grid point $(n,m)$. The values $c^2_{1;2n,2m}$
have been chosen independently and uniformly on $(-r,r)$ with $r<1/2$.
The value of $c_1^2$ is then set constant on four adjacent pixels by
enforcing that
$c^2_{1;2n-1,2m}=c^2_{1;2n-1,2m-1}=c^2_{1;2n,2m-1}=c^2_{1;2n,2m}$ for
$1\leq n,m\leq N$. In all simulations, we have $N=200$, which
generates a grid of $400^2=1.6\,10^4$ points. The time step has been
chosen so that the CFL condition $\delta t<\min\limits_{\bx}
c(\bx)/(2N)$ is ensured.

\section*{Acknowledgment}
This work was supported in part by ONR Grant
\#2002-0384. GB was supported in part by NSF Grant DMS-0072008, and LR
in part by NSF Grant DMS-9971742.

\bibliography{bibliography} \bibliographystyle{siam}

\end{document}